\documentclass{aa}
\usepackage{subcaption}
\usepackage{booktabs}

\usepackage{natbib}
\usepackage{graphicx}
\usepackage{txfonts}
\usepackage{amsmath}
\usepackage{amssymb}
\usepackage{siunitx}
\usepackage{hyperref}
\graphicspath{{Paper_final_figures/}}

\defcitealias{Bacchini_2019a}{B19a}
\defcitealias{2019b_bacchini}{B19b}
\defcitealias{2020_bacchini_SN_feedback}{B20a}
\defcitealias{Bacchini_2020_volumetric}{B20b}
\defcitealias{bacchini20243dviewlocalgravitational}{B24}

\begin{document} 
	\title{A "universal" supernova efficiency to drive the HI turbulence in nearby star-forming galaxies} 
	
	\authorrunning{L. Sassali et al.}
	
	\author{
		Lauri Sassali\inst{\ref{inst:oulu},\ref{inst:dark}},
		Cecilia Bacchini\inst{\ref{inst:dark}},
		Aku Venhola\inst{\ref{inst:oulu}}, 
		Giuliano Iorio\inst{\ref{inst:barce}},
		\and
		Antonino Marasco\inst{\ref{inst:inafpd}}
	}
	
	\institute{
		Space Physics and Astronomy Research Unit, University of Oulu, Pentti Kaiteran Katu 1, FI-90014 Oulu, Finland \label{inst:oulu}\\
		\email{lasa561@outlook.com}
		\and
		DARK, Niels Bohr Institute, University of Copenhagen, Jagtvej 155, 2200 Copenhagen, Denmark \label{inst:dark}\\
		\email{cecilia.bacchini@eso.org}
		\and
		Institut de Ciències del Cosmos (ICCUB), Universitat de Barcelona, c. Martí i Franquès 1, 08028 Barcelona, Spain \label{inst:barce}
		\and
		INAF – Padova Astronomical Observatory, Vicolo dell’Osservatorio 5, I-35122 Padova, Italy\label{inst:inafpd}\\
	}
	
	\date{}

	\abstract
	{The origin of cold gas turbulence in galaxies has long been debated. 
		Supernovae (SNe), the dominant form of stellar feedback, were thought insufficient to sustain it against dissipation, especially in galactic environments with low star formation rates (SFRs).
		We investigate the origin of the neutral atomic gas (HI) turbulence in dwarf irregular galaxies, a challenging regime of low masses and SFRs. 
		We build on a recent study of spirals showing that disc flaring prolongs the dissipation timescale, reducing the energy needed to sustain turbulence and making SN feedback viable.
		We measure the SN efficiency, that is the fraction of SN energy required to sustain HI turbulence, in 14 dwarfs, and test whether a "universal" efficiency holds across a sample representative of the star-forming main sequence.
		From the observed HI distributions and kinematics, we derive the radial profiles of the HI kinetic energy per unit area and model them as the sum of thermal energy and turbulent energy from SNe, constrained by the observed SFR surface density.
		We consider pure cold and warm neutral medium (WNM) phases as limiting cases, as well as a mixed two-phase scenario, obtaining the SN efficiency and WNM fraction via hierarchical Bayesian inference.
		We find that the observed HI kinetic energy in dwarfs requires only $\lesssim 3\%$ of the SN energy, and that a "universal" efficiency of $\approx 2\%$ explains the HI turbulence across our main-sequence sample.
		Our findings suggest that the SN efficiency remains similar across galaxies with different properties, offering valuable constraint for feedback recipes in galaxy evolution models.}

	
	\maketitle
	
	\section{Introduction}\label{sec:introduction}
	Star-forming galaxies host large amounts of cold gas in their interstellar medium (ISM), mostly in the form of neutral atomic hydrogen (HI). 
	The HI velocity dispersion, measured from the broadening of the HI 21-cm emission line, is typically $6-12 \ \mathrm{km/s}$ in star-forming dwarf galaxies \citep{Iorio_2017_data,2024Siljeg,2025ManceraPina}, slightly lower than the $8-20 \ \mathrm{km/s}$ measurements in spiral galaxies (\citet{2008_boomsma,Mogotsi_2016}, and \citet{Bacchini_2019a,2020_bacchini_SN_feedback}, hereafter \citetalias{Bacchini_2019a,2020_bacchini_SN_feedback}). 
	These velocity dispersion values are often larger than what can be explained by a purely thermal origin (i.e. $\lesssim 8 \ \text{km s}^{-1}$).  
	This observed excess in the velocity dispersions is ascribed to turbulence (i.e., chaotic motions between macroscopic portions of the fluid).
	In this context, "turbulence" indicates the line broadening produced by chaotic motions between macroscopic fluid elements along the line of sight, as opposed to the systematic expansion of shells and super-bubbles \citep[e.g.][]{Pokhrel_2020}.
	
	Numerical simulations of portions of turbulent ISM indicate that turbulence should be dissipated on timescales of order 10 Myr (e.g. \citealt{Stone_1998,Padoan_1999,MacLow_1999}). 
	Since the age of galaxies is much higher than this timescale, a continuous source is thereby required to explain the turbulence observed in the ISM. 
	Several mechanisms have been proposed for injecting the energy to sustain the turbulence.
	One of them is stellar feedback, consisting of stellar winds, ionising radiation, and supernovae (SNe), among which SNe are expected to be one of the prime drivers of turbulence \citep{Mac_Low_2004}.
	Even though, after radiative losses, only $\lesssim 10 \%$ of the energy released by SNe is left for driving turbulence (\citealt{2015_SN_energy_kim_ostriker}, but also \citealt{Fierlinger_SNR, LauBonnell2025}), SNe explode in any star-forming galaxy and release more energy than other forms of stellar feedback. 
	The physical processes governing the transition from a radially-compressive, expanding shell into a fully developed turbulent flow remain an open area of research and include hydrodynamical instabilities, shocks, and other complex processes \citep[see][]{Vishniac1983,LauBonnell2025,Beattie_2026}.
	
	A few observational studies investigating the origin of turbulence find that SNe contribute but cannot fully account for the observed turbulent energies \citep[e.g.][]{Stilp_2013_drivers_of_turbulence,Tamburro_2009}. Therefore, mechanisms other than stellar feedback, such as local gravitational instabilities, gas accretion, and magneto-rotational instabilities (MRI) have been proposed to explain the observed turbulence \citep{Mac_Low_2004}, but no general consensus was reached on the prime driver of turbulence.
	The study by \citetalias{2020_bacchini_SN_feedback} proposed a solution to this conundrum by considering that the flaring of the gas disc prolongs the dissipation timescale, so the energy needed to sustain the turbulence is reduced. 
	Thus, they find that only a small fraction ($\approx 1.5\%$) of the total SN energy is sufficient to maintain turbulence in the atomic gas of spiral galaxies. 
	The efficiencies obtained by  \citetalias{2020_bacchini_SN_feedback} are well within the 0.1 limit imposed by SN remnant models \citep{2015_SN_energy_kim_ostriker} and also leave some energy to drive other galactic phenomena, e.g. ionised gas winds or galactic fountains  \citep{2017MarascoFraternali,2023Marasco}. 
	
	However, \citetalias{2020_bacchini_SN_feedback} is limited to galaxies with relatively high stellar masses and SFRs, leaving it unclear whether SNe are sufficient to sustain the HI turbulence of dwarfs. 
	Even though the HI turbulence in dwarfs is just a factor $\lesssim 2$ lower than in spirals, the star formation rates (SFRs) in dwarfs can be $\lesssim 1-2$ orders of magnitude lower. 
	Therefore, dwarfs offer compelling case studies to study the origin of turbulence in extreme conditions. 
	The efforts so far, although with important differences in the adopted methodologies, have led to the conclusions that SNe alone are insufficient to explain the HI turbulence in dwarfs. 
	Besides SNe, \citet{Stilp_2013_drivers_of_turbulence} considered MRI, gravitational instabilities, and accretion as possible contributors. However, they concluded that none of the mechanisms could fully account for the observed HI turbulence in dwarfs. 
	On the other hand, \citet{Hunter_2021_correlation} propose that mild local gravitational instabilities are the most likely option. 
	However, a few empirical studies have shown that HI discs in dwarfs are not susceptible to this type of instability \citep{2008Begum_a,2012Elson,2017Namumba,2018Watts,bacchini20243dviewlocalgravitational}. 
	Therefore, the origin of turbulence in dwarf galaxies remains poorly understood. 
	
	Unravelling the physical mechanisms driving the HI turbulence in galaxies is complicated by two difficulties. 
	The first is the multiphase nature of HI. 
	The HI is thought to be mostly distributed into two phases, the cold neutral medium (CNM) and the warm neutral medium (WNM) (\cite{1977_McKee_Ostriker}, \citet{1987_kulkarni_heiles}). 
	Constraining their relative abundances is a challenging task that is viable only for the Milky Way and very nearby galaxies. 
	A few studies so far indicate that there is more WNM than CNM. In the solar neighbourhood, around $60\%$ of the HI is WNM \citep[e.g.][]{2003_Heiles_Troland_WNM}, while \citet{2013_pineda} report $\sim 30\%$ and $\sim 80\%$ within and beyond the solar radius, respectively, perhaps hinting at an increasing trend with the Galactocentric distance.
	In dwarf galaxies, the available estimates are similarly high, ranging from $65 \%$ for the Large Magellanic Cloud \citep{2000-zimmer} to $\sim 85\%$ for the Small Magellanic Cloud \citep[][see also \citealt{Young_2003_WNM_dIrrs}]{Dickey_2000}.
	This is thought to be a consequence of the low metallicity and the diffuse nature of the ISM in dwarfs \citep{hunter2024interstellarmediumdwarfirregular}: at low metallicities ($\approx 10^{-3}$--$10^{-2} \, Z_{\odot}$) radiative cooling becomes increasingly inefficient due to the reduced abundance of dust, metals, and primary coolant molecules such as $\text{C}^+$, O and CO, while cosmic ray heating remains efficient even where photoelectric heating is not.
	Moreover, the shallow potential wells of dwarfs make their ISM diffuse and set low midplane pressures, at which the CNM is no longer thermally stable and the WNM becomes the dominant phase (e.g. \citealt{wolfire_2003_temperatures}, \citealt{2014_Saury}).
	
	The second challenge in understanding the origin of turbulence is that different mechanisms can contribute, each with its own efficiency. 
	Estimating their relative roles is further complicated by observational uncertainties. MRI, for instance, depends on the magnetic field strength, which is difficult to constrain and varies across a galaxy \citep{1996Beck,2008Chyzy}, and its efficiency is essentially unconstrained; gas accretion and gravitational instabilities suffer from similar issues. SN feedback is less problematic in this respect, as it is constrained by the observed SFR distribution and by the efficiencies $\lesssim 10$\% predicted by SN remnant models.
	
	The aim of this study is twofold. 
	First, we want to clarify the role of SN feedback in sustaining the observed HI turbulence, complementing the study \citetalias{2020_bacchini_SN_feedback} by considering a sample of nearby dwarfs with very low mass and SFRs. Second, we aim to understand whether a "universal" SN efficiency for star-forming galaxies exists or different SN efficiencies are required for dwarfs and spirals \citep{Hunter_2021_correlation}. 
	This is essential to clarify how stellar feedback impacts the ISM and how its energy budget is spent. 
	A "universal" SN efficiency would also be particularly useful for numerical simulations of galaxies, which have to adopt a given efficiency in their sub-grid prescriptions for stellar feedback. 
	Furthermore, we want to investigate whether the fraction of WNM can be constrained under the assumption that the SN efficiency must be the same for any galaxy. 
	
	This paper is structured as follows. Section \ref{sec:sample} presents the galaxy sample, the data, and the derivation of the HI surface density, scale height, and SFR surface density. Section \ref{sect:methods} describes the model and the hierarchical Bayesian framework. Section \ref{sec:results} reports the results for the different HI phase assumptions, both for individual galaxies and for a "universal" efficiency. Section \ref{sect:discussion} discusses the implications, the comparison with the literature, and the caveats, and Section \ref{sect:conclusion} summarises our findings.
	
	\section{Galaxy sample and datasets}\label{sec:sample} 
	The SN efficiency is constrained using the radial profile of gas kinetic energy per unit area derived from observables as (\citealt{Tamburro_2009}, \citetalias{2020_bacchini_SN_feedback})
	\begin{equation}\label{eq:obs_E}
		E_{\text{obs}}(R) = \frac{3}{2}\Sigma(R)\sigma^2(R),
	\end{equation}
	where $\Sigma(R)$ is the gas surface density and $\sigma(R)$ is the gas velocity dispersion. 
	Then, the kinetic energy released by SN feedback across the disc is estimated from observations using the radial profile of the SFR surface density ($\Sigma_\mathrm{SFR}$; see Sect.~\ref{sec:modelling}). 
	Therefore, we selected the sample of dwarfs based on the availability of robust measurements of the HI distribution and kinematics, and surface photometry from FUV and NUV images.  A sub-sample of nearby 14 dwarfs from the Local Irregulars That Trace Luminosity Extremes, The HI Nearby Galaxy Survey (LITTLE THINGS; \citet{2012_LITTLETHINGS}) meets these requirements. Table \ref{tab:galaxy_data} summarises the properties of the galaxy sample. We note that two galaxies in our sample, DDO~50 and NGC~2366, are classified as dwarf starbursts.
	
\begin{figure}[ht]
		\centering    \includegraphics[width=1\linewidth]{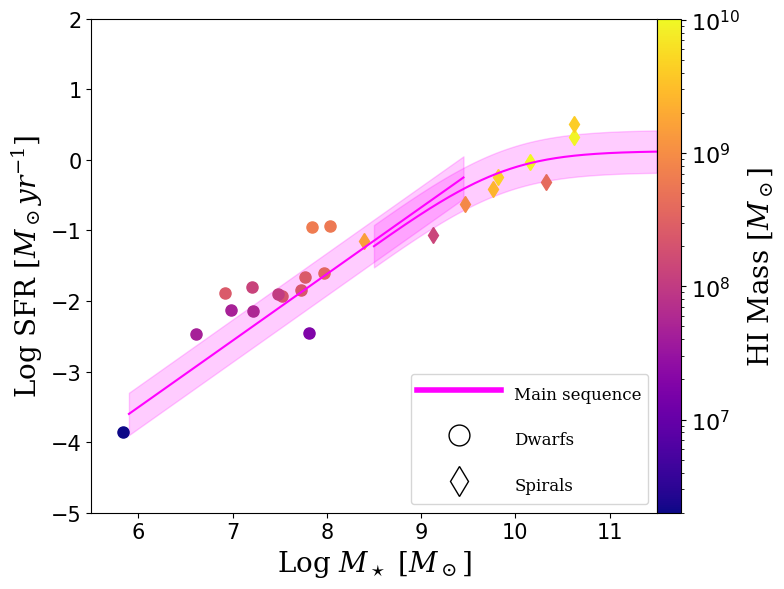}
		\caption{Stellar mass versus SFR for our sample of dwarf (circles) and spiral (diamonds) galaxies. Solid curves are the star-forming main sequence for galaxies with $6\lesssim \log(M_\star / M_\odot) \lesssim 9.5$ \citep{Berg2022} and $8.5 \lesssim \log(M_\star / M_\odot) \lesssim 11.5$ \citep{2023_Popesso}, with the shaded bands showing the 0.3 dex scatter.}
		\label{fig:MS_distribution}
	\end{figure}

	We also include the sample of nearby spiral galaxies used in \cite{2020_bacchini_SN_feedback}, which covers a range of stellar masses higher than our sample. 
	Figure~\ref{fig:MS_distribution} shows that, combined together, our samples of dwarfs and spirals are representative of the star-forming main sequence for present-day galaxies. 

	\subsection{HI surface density and velocity dispersion}\label{sect:kinematics}
	The radial profiles of the HI surface density and velocity dispersion were obtained by \citet{Iorio_2017_data}, leveraging 21-cm emission line data cubes from  LITTLE THINGS \citep{2012_LITTLETHINGS}.  
	The data cubes were modelled with the publicly available software $^\mathrm{3D}$BAROLO \citep{2015_barolo}, which fits a tilted-ring model to emission line data cubes to measure the gas distribution and kinematics. This is a first improvement with respect to previous studies of the HI turbulence in dwarfs \citep{Hunter_2021_correlation,Stilp_2013_drivers_of_turbulence}, since $^\mathrm{3D}$BAROLO corrects for beam smearing, which artificially increases the velocity dispersion measured from low-resolution data. Those studies rely instead on pixel-by-pixel measurements of $\sigma_\mathrm{HI}$, which are affected by beam smearing, or on stacking line profiles across the disc, which may add artificial broadening if the profiles are not perfectly aligned. For a few dwarfs, \cite{Iorio_2017_data} note local discontinuities in the $\sigma_\mathrm{HI}$ profiles, typically in regions of low signal-to-noise ratio (e.g. the HI disc outskirts, cavities, shells) or with kinematic anomalies (e.g. non-circular motions). We exclude those data points, which are only $\approx 5\%$ of the total and do not affect our conclusions.
	For DDO~101, we use updated measurements of HI distribution and kinematics, obtained by \cite{2022ManceraPina_b} assuming an improved distance measurement.
	
	The radial profiles of the HI surface density and velocity dispersion for the sample of spirals from \cite{2020_bacchini_SN_feedback} were also measured using $^\mathrm{3D}$BAROLO to model 21-cm emission line data cubes from the surveys "The HI Nearby Galaxy Survey" (THINGS; \cite{THINGS_2008}) and "Hydrogen Accretion in LOcal GAlaxieS (HALOGAS) Survey \citep{2011_Heald_HALOGAS}.

	\subsection{HI disc scale heights}\label{sec:mass_models}
	The dissipation timescale of turbulence in a gas disc depends on the scale height of the gas disc itself. 	
	For both spiral and dwarf galaxies, the vertical distribution of HI and the corresponding HI scale height ($h_\mathrm{HI}$) are calculated numerically, assuming that the HI disc is in vertical hydrostatic equilibrium within the galactic potential. 
	This procedure, detailed in \cite{Bacchini_2019a}, is based on an iterative algorithm that, given the gas velocity dispersion measured from observations and a mass model for the galaxy, calculates the vertical distribution of a gas disc taking into account its self-gravity and gives the gas scale height as a function of the galactocentric distance. The scale height is the standard deviation of the vertical density distribution (Eq. (1) in \citetalias{Bacchini_2020_volumetric}). We notice that this approach implicitly assumes that the scale height is the same for all HI phases, while in principle the CNM and WNM might have different scale heights (as CO and HI, for instance; see \citealt{2017Marasco} and \citetalias{Bacchini_2019a}).  
	This is equivalent to assuming that the CNM and WNM have the same velocity dispersion, but with different contributions from thermal and turbulence motions (see Sect.~\ref{sec:modelling}).
	
\begin{figure}[ht]
		\centering
		\includegraphics[width=1.\linewidth]{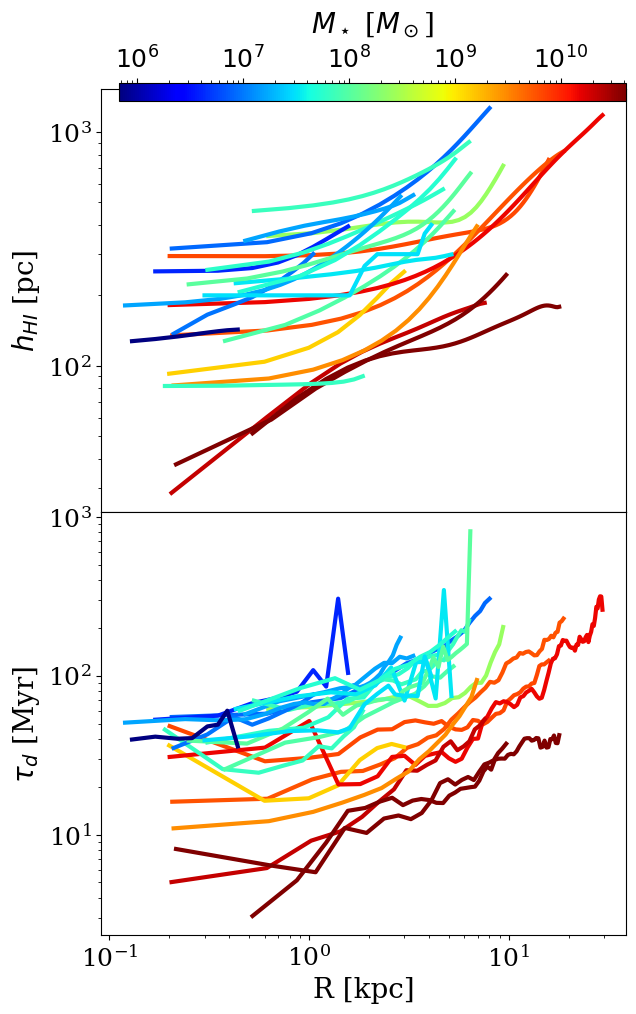}
		\caption{Radial profiles of the HI scale height (top) and the dissipation timescale of HI turbulence (bottom) for dwarfs and spirals. The profiles are coloured according to the galaxy stellar mass. The dissipation timescale is calculated using Eq.~\ref{eq:tau} and $\sigma_\mathrm{HI}$.}
		\label{fig:DissipationTime}
	\end{figure}

	For spirals, the radial profiles of $h_\mathrm{HI}$ were derived in \citetalias{Bacchini_2019a}, while those for dwarfs are from \cite{Bacchini_2020_volumetric} (hereafter \citetalias{Bacchini_2020_volumetric}) and \citetalias{bacchini20243dviewlocalgravitational}. 
	The top panel in Fig.~\ref{fig:DissipationTime} displays the radial profiles of $h_\mathrm{HI}$ for the samples of spirals and dwarfs analysed in this work. 
	Overall, $h_\mathrm{HI}$ tends to increase with decreasing $M_\star$. 
	However, the low-mass galaxies do not necessarily have thicker HI discs than high-mass ones, because $h_\mathrm{HI}$ depends more on the interplay between the $\sigma_\mathrm{HI}$ and gravitational potential. 
	We notice that $h_\mathrm{HI}$ also correlates with the halo mass, as expected since the two correlate and the dark matter dominates the potential in dwarfs and spiral outskirts \citep[see][]{2025ManceraPina}. 
	
	The mass models for dwarf and spiral galaxies were derived via the rotation curve decomposition into the contribution of the different mass components, including the atomic gas disc, a molecular gas disc (when present in spirals), a stellar disc, a bulge (when present in spirals), and a dark matter (DM) halo. 
	For dwarfs, the mass models were derived by \cite{2016_read_DM_core} using the HI circular velocity profiles derived by \cite{Iorio_2017_data}. 
	This latter study includes a robust estimate for the asymmetric drift correction to the HI rotation curve, crucial in the case of dwarfs where pressure support can be significant. 
	In the case of dwarfs, the DM halo is modelled as a cored Navarro-Frenk-White (NFW) profile \citep{1995_navarro_og, 1996_navarro_cored}. 
	This kind of DM density profile was obtained from high-resolution numerical simulations of dwarfs, in which the central density (or concentration) of the cusped NFW halo gradually decreases due to the energy and momentum released by stellar feedback (see also \citet{2005_read_gilmore}; \citet{2014_Pontzen_Governato}).  
	For spiral galaxies, the mass models are taken from \citet{de_Blok_2008} or \citet{2016_frank} (for details see also \citetalias{Bacchini_2019a}).
	
	\subsection{SFR surface density}
	Surface photometry for the sample of dwarf galaxies was obtained by \citet{Hunter_2010_data,zhang2012_stellarmass} using archival UV images from the Galaxy Evolution Explorer (GALEX). Each galaxy was imaged simultaneously in two bandpasses, FUV ($1350 - 1750~\AA$) and NUV $(1750 - 2800~\AA$), with an angular resolution of $5.6^{\prime\prime}$ and $4.0^{\prime\prime}$, respectively. To derive radial profiles of UV surface brightness, each galaxy image was divided into concentric elliptical annuli, with fixed centre positions, PAs, and ellipticities. We note that the ellipticities and PAs used to derive the UV surface brightness radial profiles are not necessarily the same used for the HI kinematic modelling (Sect. \ref{sect:kinematics}), as the geometry of the HI disc might differ from that of young stellar populations, especially for dwarf irregulars. 
	
	UV-photometry was corrected for both Galactic foreground reddening and internal reddening. Foreground $E(B-V)$ values were adopted from \citet{1984_u_corr}, while a uniform internal reddening of $E(B-V) = 0.05 \ \text{mag}$ was assumed, following typical values for dwarfs. UV extinction corrections were then applied using the extinction law of \citet{2007_Wyder} extinctions  $A_{\text{FUV}} = 8.24 E(B-V)_t$ and $A_{\text{NUV}} = 8.20E(B-V)_t$. A single extinction correction was applied across all radii within each dwarf, justified by typically low metallicity and dust gradients in dwarfs. To convert the UV photometric data to SFR surface density, we follow the same procedure as \citetalias{2020_bacchini_SN_feedback} and assume the relation between FUV luminosity $L_{\nu}\mathrm{(FUV)}$ and SFR from \cite{McQuinn_2015} and derived specifically for dwarfs
	\begin{equation}\label{eq:sfr}
		\mathrm{SFR} = (1.37 \pm 0.54) \times 10^{-28} \left [\frac{L_{\nu}\mathrm{(FUV)}}{\text{erg} \ \text{s}^{-1} \text{Hz}^{-1}} \right ] M_\odot \text{yr}^{-1} \, ,
	\end{equation}
	where the normalisation has been multiplied by 0.67 to rescale it from a Salpeter to a Kroupa IMF. 
	
	The UV emission often extends up to smaller galactocentric distances than the HI distribution, as the HI can still be detected in the outer regions. 
	This does not necessarily mean that no star formation is occurring in those outer regions, but it could be due to the fact that the UV surface brightness drops below the detection limit. 
	For this reason, we decided to assign an upper limit to $\Sigma_{\text{SFR}}$ in the regions where HI is detected while UV is not.
	We estimated upper limits on $\Sigma_\text{SFR}$ by extrapolating the observed UV surface brightness profile.
	We fit the outermost five data points of $\log \Sigma_{\text{SFR}}$ with a linear model (as $\Sigma_{\text{SFR}}(R)$ typically follows an exponential decline). 
	The upper limit is given by the value of the best-fit model evaluated at the radius immediately beyond the outermost measurement of the observed UV surface brightness profile (see Sect.~\ref{sect:discussion} for discussion). 
	
	For the galaxies from \citetalias{2020_bacchini_SN_feedback}, the SFR surface densities are derived in a slightly different way than for dwarfs. Specifically, the SFR surface densities are taken from \citet{Leroy_2008}, who also assumed a Kroupa IMF but used a combination of FUV and 24 $\mu$m luminosities to take into account the emission from massive and young stars absorbed by dust and re-radiated at the mid-infrared wavelengths. As pointed out in \citetalias{Bacchini_2020_volumetric}, the SFR-luminosity relation used by \citet{Leroy_2008} was obtained for normal star-forming galaxies and may not be applicable to dwarf galaxies \citep{Hao_2011}.

	\section{Methods}\label{sect:methods}
	
	\subsection{Modelling of the atomic gas energy}\label{sec:modelling}
	The observed kinetic energy is modelled as the sum of the turbulent ($E_{\mathrm{turb}}$) and thermal ($E_{\mathrm{th}}$) energy components (\citealt{Tamburro_2009}, \citetalias{2020_bacchini_SN_feedback})
	\begin{equation}\label{eq:E_mod}
		E_{\text{mod}}(R) = E_{\mathrm{turb}}(R) + E_{\mathrm{th}}(R) \, .
	\end{equation}
	Thus, the total velocity dispersion of the gas we observe is
	\begin{equation}\label{eq:sigma_mod}
		\sigma_{\text{mod}}(R) = \sqrt{v^2_{\text{turb}}(R)+v^2_{\text{th}}(R)} \, ,
	\end{equation}
	with $v_{\text{turb}}$ the turbulent velocity and $v_{\text{th}}$ the thermal velocity.
	
	The timescale of turbulence dissipation ($\tau_\mathrm{d}$) is defined as the ratio between the turbulent energy ($E_\mathrm{turb}$) and its dissipation rate ($\dot{E}_{\mathrm{diss}}$). 
	The gas turbulence can only be maintained if fresh energy is injected at the driving scale $L_\mathrm{D}$, at the same rate as energy is transferred to the smallest scales, hence $\mathrm{\dot{E}_{\mathrm{diss}}} = \dot{E}_\mathrm{turb,SNe}$ with $\dot{E}_\mathrm{turb,SNe}$ being the SN energy injection. Therefore, the timescale for dissipating a given amount of turbulent energy is
	\begin{equation}\label{eq:tau}
		\tau_\mathrm{d} \equiv \frac{E_{\text{turb}}}{\dot{E}_{\text{diss}}} = \frac{E_{\text{turb}}}{\dot{E}_{\text{turb}}} = \frac{L_D}{v_\text{turb}} \, ,
	\end{equation}
	which shows $\tau_\mathrm{d}$ corresponds to the crossing time of turbulent gas across the driving scale \citep{2000_turbulence_crossingtime_elmegreen,Stone_1998}. 
	We have implicitly assumed that the injection scale of turbulence in Eq.~\ref{eq:tau} is $L_D = 2h_\mathrm{HI}$. There is indeed observational evidence that the explosion of multiple SNe, in combination with other forms of stellar feedback (e.g. winds), generate expanding shells called super-bubbles, whose diameter can easily reach the thickness of the HI disc (i.e. $2h_\mathrm{HI}$; see for example \citealt{2008_boomsma}, \citealt{1989_SNR_blowout_maclow}). 
	The bottom panel in Fig.~\ref{fig:DissipationTime} shows the radial profiles of $\tau_\mathrm{d}$ for dwarfs and spirals, using $v_\text{turb} = \sigma_\mathrm{HI}$ in Eq.~\ref{eq:tau}. 
	For all galaxies, $\tau_\mathrm{d}$ increases with the galactocentric distance, becoming even one order of magnitude longer in the HI disc outskirts ($\tau_\mathrm{d} \sim 100$~Myr) than in the inner regions ($\tau_\mathrm{d} \sim 10$~Myr). 
	
	\subsubsection{Turbulent energy from SNe}
	Given the observed $\Sigma_\mathrm{SFR}$, the rate of kinetic energy injection by multiple SN explosions per unit area and dissipation timescale can be calculated as \citep{Tamburro_2009}\citetalias{2020_bacchini_SN_feedback}
	\begin{equation}\label{eq:EturbSN}
		\dot{E}_\mathrm{turb,SNe} = \eta \Sigma_\mathrm{SFR} f_\mathrm{cc} E_\mathrm{SN},
	\end{equation}
	where $\eta$ is the dimensionless SN energy coupling efficiency, $f_\mathrm{cc} \approx 1.3 \times 10^{-2} \ M^{-1}_\odot$ is the number of core-collapse SNe that explode for the unit of stellar mass formed, and $E_\mathrm{SN} = 10^{51}$~erg is the total energy released by a single SN \citep{2015_SN_energy_kim_ostriker}.
	The value of $f_\mathrm{cc}$ is obtained by integrating the initial mass function (IMF) over the mass range of stars that end up in the evolution as core collapse SNe. We assume a Kroupa (2002) IMF and a mass range of $8 \ M_\odot \ - \ 120 \ M_\odot$ (see, e.g. \citealt{1931_chandrasekhar,2009_stellar_ev}).
	The turbulent energy injected by SNe over a dissipation timescale is obtained by combining Eqs. \ref{eq:tau} and \ref{eq:EturbSN}:
	\begin{equation}\label{eq:SN_E_1}
		E_\mathrm{turb,SNe} = \eta\Sigma_\mathrm{SFR}f_\mathrm{cc}E_\mathrm{SN}\frac{2h}{v_\mathrm{turb}}, 
	\end{equation}
	We note that $E_\mathrm{turb,SNe}$ is nearly independent from $\sigma$, since $h$ from the hydrostatic equilibrium is proportional to $\sigma$ (Sect.~\ref{sec:mass_models}).
	
	\subsubsection{Thermal energy contribution}
	The neutral atomic gas in the ISM consists of WNM and the CNM. 
	The contribution of these two phases to the observed kinetic energy depends on their temperatures and mass fractions. 
	In fact, the gas thermal velocity at a given temperature $T$ is $v_\text{th} \equiv \sqrt{k_BT/(\mu m_p)}$, where $\mu \approx 1$ is the mean particle weight in units of proton mass $m_p$ for HI and $k_B$ is the Boltzmann constant.
	The thermal energy per unit area is 
	\begin{equation}\label{eq:E_th}
		E_\text{th}(R) = \frac{3}{2}\Sigma(R)\frac{k_BT}{\mu m_p}
	\end{equation}
	Hence, the total thermal energy per unit area, including the contribution of both CNM and WNM is
	\begin{equation}\label{eq:Eth_final}
		E_\text{th} = E_\text{CNM} + E_\text{WNM} = \frac{3}{2}\Sigma\left [(1-f_\mathrm{w})v_\mathrm{th,c}^2+f_\mathrm{w} v_\mathrm{th,w}^2\right ] \, ,
	\end{equation}
	where $f_\mathrm{w}$ is the WNM-to-total HI fraction, and $v_\mathrm{th,c}$,$v_\mathrm{th,w}$ are the thermal velocities of CNM and WNM, respectively. 
	For consistency with \citet{2020_bacchini_SN_feedback}, we adopt the temperature ranges of $T \approx 40 - 190$ K for the CNM and $T \approx 7000 - 8830$ K for the WNM, motivated by the average temperatures resulting from the model by \citet{wolfire_2003_temperatures}. However, the observed temperatures in ISM show dependency on metallicity, i.e. in low-metallicity environments such as dwarfs, the temperatures may change e.g., $\sim 30$ K for the CNM in Small Magellanic Cloud (see e.g. \citealt{Jameson_2019}, \citealt{Chen_2025}). However, this small difference in $T$ does not impact our final conclusions, given the low contribution of CNM to the total thermal energies and that $f_\mathrm{w} \propto T^{1/2}$.
	These temperature ranges correspond to thermal velocities of $0.6 \ \text{km s}^{-1}\lesssim v_\text{th,c} \lesssim 1.3 \ \text{km s}^{-1}$ for the CNM and  $7.6 \ \text{km s}^{-1}\lesssim v_\text{th,w} \lesssim 8.6 \ \text{km s}^{-1}$ for the WNM. Accordingly, our model assumes distributions of $v_\mathrm{th}$ centred on the mean values of these intervals, namely $1.0 \ \text{km/s}$ for the CNM and $8.1 \ \text{km/s}$ for the WNM, with dispersions of $0.5 \ \text{km/s}$ to cover the range of plausible values (see Sect~\ref{sect:methods}). This choice implicitly assumes the HI gas is thermally stable, whereas observations suggest that part of the HI may be thermally unstable, at least in the Milky Way \citep{2003_Heiles_Troland_WNM} (see also \citealt{Roy_2013}, \citealt{Kalberla_2018}). This assumption does not impact our conclusions regarding the SN efficiency. 
	However, our interpretation of $f_\mathrm{w}$ implicitly assumes thermal stability so, for completeness, the online material also provides the results in terms of $v_\mathrm{th}$, which can be converted into $T$ by inverting $v_\mathrm{th}$ definition without assuming thermal stability.
	
	\subsection{Hierarchical Bayesian inference}\label{sec:bayesian_method}
	We adopt the same hierarchical Bayesian approach of \citetalias{2020_bacchini_SN_feedback}, which includes a detailed description of the probability functions and methods used here (see especially Sect.~3.2 and Appendix~B). This approach ensures robust treatment of uncertainties and upper limits for both the observed quantities and the model variables. The hierarchical model is implemented via the PyMC3 Python package using the Hamiltonian Monte Carlo sampling.  We apply the same algorithm to all three cases of interest: i) $100\%$ WNM, ii) $100\%$ CNM, and iii) two-phase medium partially consisting of both CNM and WNM. 
	In the following, we provide a brief overview of this approach. 
	
	Hierarchical Bayesian models introduce a further level of variability to the classical Bayes' theorem, as the model parameters are described with an additional set of hyperparameters. The Bayes' law thus takes the form of
	\begin{equation}
		P(\theta\mid D) \propto P(D \mid \theta)  P(\theta \mid \phi) P(\phi),
	\end{equation}
	where $P(\theta \mid D)$ is describing the probability distribution of the model parameters $\theta$ given the observed data $D$, $P(D\mid \theta)$ is a conditional probability of $D$ given $\theta$ (i.e. the likelihood of $\theta$ given $D$), $P(\theta \mid \phi)$ is the probability distribution of the priors given the hyperpriors $\phi$, and $P(\phi)$ is the hyper-prior. 
	
	In practice, the observed quantities ($\Sigma_\mathrm{SFR}$, $\sigma$, $h_\mathrm{HI}$, $\Sigma_\mathrm{HI}$) are modelled as realisations of normal distributions centred on unknown true values ($\Sigma_\mathrm{SFR}^T$, $\sigma^T$, $h_\mathrm{HI}^T$, $\Sigma_\mathrm{HI}^T$), with standard deviations set by the observational uncertainties and log-normal priors. The comparison between measured and predicted true values defines the likelihood, which drives the sampling towards the most probable realisations. 
	For the sampling, we ran four independent chains with 20,000 sampling steps each, discarding the initial 5,000 samples as burn-in. Convergence was assessed using trace plots and the $\hat{R}$ statistic, which compares the variance between multiple chains to the variance within each chain \footnote{For a more detailed description see \url{https://python.arviz.org/en/v0.23.4/api/generated/arviz.rhat.html}.}. 
	
	We determine the free parameters of interest by marginalising over all other parameters in the hierarchical model. In case iii), we have both the SN feedback efficiency ($\eta$) and the WNM gas fraction ($f_\mathrm{w}$), each described by its own set of probability distributions (see \citetalias{2020_bacchini_SN_feedback} for details). These parameters are partially degenerate, although the observed velocity dispersions might constrain $f_\mathrm{w}$ in regions with low $\Sigma_\mathrm{SFR}$ (i.e. where SN feedback is expected to be weak). In cases i) and ii), we consider only the SN feedback, and the $f_\mathrm{w}$ takes values 0 or 1, respectively.
	
	The observational quantities $\Sigma$, $\Sigma_\mathrm{SFR}$, and $\sigma$ exhibit large uncertainties, mostly because of the low surface brightness of HI and UV, the asymmetric HI distribution, and the presence of non-circular motions that complicate the HI kinematics \citep[for a detailed discussion, see][]{Iorio_2017_data}. These uncertainties can become especially large in the outer regions of galactic discs. In a hierarchical model, such large uncertainties weaken the constraints on the true values, which in turn can result in inefficient sampling and poor mixing. To mitigate this issue, we adopt a non-centred parametrisation. This approach has been shown to improve sampling efficiency when the true quantities are weakly informed by the data and to alleviate pathological posterior geometries, often described as “funnel”-shaped structures (see, e.g., \citealt{papaspiliopoulos}, \citealt{betancourt2013hamiltonianmontecarlohierarchical}). These geometries occur when the hyperprior controlling the variability of the model parameters ($\theta$) approach very small values, creating regions of extremely high curvature that are correspondingly slow to explore. In practice, we introduce an additional “offset” hyperparameter and rewrite the log-normal distributions of the true values through a non-centred parametrisation. In PyMC3, this effectively turns the true values into deterministic nodes (i.e. defined purely by their parent nodes, without any add of randomness) and improves the posterior geometry, thereby enabling more efficient sampling. 
	
	\section{Results}\label{sec:results}
	In this section, we present the results of this study and their interpretation, concentrating on the inferred values of SN efficiency needed to maintain turbulence in the atomic gas.
	The analysis proceeds in steps of increasing realism and generality. We first derive characteristic SN efficiencies for each dwarf individually, considering three limiting assumptions about the HI temperature distribution: a purely cold medium (100\% CNM, Sect.~\ref{sec:cnm}), a purely warm medium (100\% WNM, Sect.~\ref{sec:wnm}), and a more realistic two-phase gas (Sect.~\ref{sec:2phase_single}). The CNM and WNM cases are directly comparable to the characteristic SN efficiencies for spirals in \citetalias{2020_bacchini_SN_feedback}. Building on this, we then test the hypothesis of a single “universal” SN efficiency. In practice, in our modelling, we impose that the SN efficiency does not vary between galaxies. 
	We first test whether a single efficiency can simultaneously reproduce the HI turbulence across all dwarfs (Sect.~\ref{sec:GlobalEta_separated}), and finally across the full star-forming main sequence by combining dwarfs and spirals (Sect.~\ref{sec:GlobalEta}). Each step is motivated by the results of the previous one, and the universal efficiency in the two-phase case for the combined sample is our headline result.
	
	\subsection{Characteristic SN efficiencies of dwarf galaxies}\label{sec:single_results}
	
	\subsubsection{Cold atomic gas}\label{sec:cnm}
	When all atomic gas is in the CNM phase (i.e. $f_\mathrm{w}$ = 0), thermal motions give a minimum contribution to the total energy budget. Therefore, the obtained efficiency ($\eta_\mathrm{CNM}$) can be interpreted as an upper limit (under the assumption of negligible radiative losses, see Sect.~\ref{sect:discussion} for discussion). 
	Figure \ref{fig:Energycomponents-CNM-single} shows the contribution of the different energy components to the total energy budget for each of the galaxies in our sample. 
	The SN energies play a dominant role in most radii, but given that $\Sigma_\mathrm{SFR}$ decreases with radius, this component generally becomes less important at larger radii and can even be comparable to the thermal energy in some cases. Our models reproduce the overall trend of the observed energies, but they do not strictly follow them either. 
	The largest discrepancy can be seen in DDO~154, DDO~52 and WLM at the radii where we assume the upper limit for $\Sigma_\mathrm{SFR}$.
	
\begin{figure*}[ht]
		\centering   \includegraphics[width=1.\linewidth]{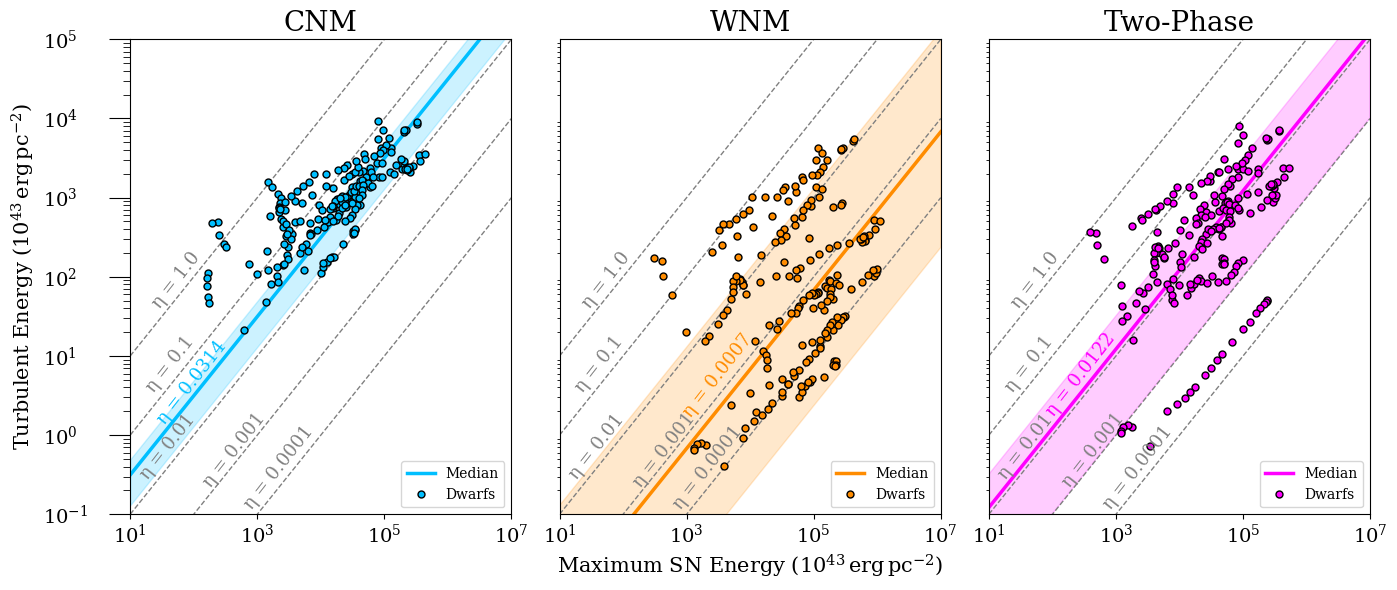}
		\caption{Maximum energy per unit area provided by SNe in a dissipation timescale (i.e. for $\eta = 1$) versus the turbulent component of the observed energy per unit area. Each point is for a single radius for a single galaxy. The panels show the three cases: 100\% CNM (left panel) and 100\% WNM (central panel) and the two-phase case (right panel). Grey dotted lines indicate relations $E_\mathrm{obs,turb} = \eta E_\mathrm{turb, SNe}$ with $\eta = 10^n$ and $n = -4, -3... , 0$. The solid lines indicate the median efficiency with the shaded areas outlining the $1\sigma$ uncertainties when considering all posterior distributions of dwarfs in the respective cases. For clarity of the figure, the uncertainties are not displayed here.}
		\label{fig:All_IN}
	\end{figure*}

	Table~\ref{tab:SingleEffs} reports the characteristic SN efficiencies for each dwarf and their median value, which is calculated by extracting $10^4$ random values from all the posteriors. 
	The efficiencies are $\eta_\mathrm{CNM} \lesssim 0.09$ with a median $\langle \eta_\mathrm{CNM}\rangle \approx 0.031$. 
	Figure~\ref{fig:All_IN} shows the radial profiles of the turbulent energy component versus the maximum energy from SNe. 
	While most measurements are consistent with the median efficiency within the uncertainties, it is clear that a significant number of data points require efficiencies above 0.1 or even 1. 
	However, a scenario in which all the HI is CNM is not realistic and likely underestimates the thermal energy contribution.
	
	\subsubsection{Warm atomic gas}\label{sec:wnm}
	In the 100\% WNM case, the contribution from thermal motions is maximised. 
	Therefore, the efficiencies ($\eta_\mathrm{WNM}$) obtained are considered to be a lower limit. 
	Figure~\ref{fig:Energycomponents-WNM-single} shows the radial profiles of each energy component for our galaxies. The SN energies become less important than in the 100\% CNM case. 
	The model reproduces the observed energies better than in the CNM case, underlining the importance of the WNM thermal energy. However, some galaxies require very low efficiencies ($\sim 10^{-4} - 10^{-5}$) and the model even slightly exceeds the observed energy, suggesting that a WNM fraction of 100\% is too high: $E_\mathrm{turb, SNe}$ dominates in the inner regions of a few galaxies, but is generally not needed in the outskirts.
	
	The characteristic efficiencies obtained are $\eta_\mathrm{WNM} \lesssim 0.03$ with median $\langle \eta_\mathrm{WNM}\rangle \approx 0.001$ (see Table~\ref{tab:SingleEffs}). 
	These values are $\approx 3$ times lower than the characteristic SN efficiencies for the spirals in \citetalias{2020_bacchini_SN_feedback}. 
	In Fig.~\ref{fig:All_IN}, the central panel shows the radial profiles of the turbulent energy and the maximum energy from SNe. With respect to the CNM case, the scatter between the points is much larger, spanning almost five orders of magnitude. 
	This suggests that also this extreme and unrealistic case in which 100\% of the HI is WNM, is not the best to reproduce the data. 
	
	\subsubsection{Two-phase atomic gas}\label{sec:2phase_single}
	Finally, we consider the most realistic scenario in which the HI consists of both CNM and WNM. We use the same methodology as in the previous two cases, but the thermal energy is calculated using Eq.~\ref{eq:Eth_final}. The thermal speed is now treated as the second free parameter, allowing us to obtain the fraction of WNM ($f_\mathrm{w}$) along with the SN efficiency ($\eta_\mathrm{2ph}$). This choice yields a more realistic estimate of the uncertainties, because the $\eta_\mathrm{2ph}$ posterior is marginalised over $f_\mathrm{w}$. 
	
	Figure~\ref{fig:Energycomponents-2phase-single} shows that the observed energies are reproduced very well in most galaxies, better than in the extreme cases discussed in Sects.~\ref{sec:cnm} and~\ref{sec:wnm}.  
	Both the total thermal energy component and the SN component are important in these galaxies. The energy from SNe is again dominant mostly in the inner regions, but it contributes less in the outer regions. However, it seems that in dwarf galaxies the difference between these two components is less apparent than in spiral galaxies (see Fig.~3 from \citetalias{2020_bacchini_SN_feedback}). 
	Even though the second free parameter improves the match between model and observations, the skewed posteriors (see atlas in online material) show that we cannot fully break the degeneracy between $f_w$ and $\eta$, i.e. we cannot disentangle thermal motions from turbulence, likely because of the large observational uncertainties. Both parameters are therefore less well constrained than in the 100\% CNM and 100\% WNM cases.
	In Section~\ref{sect:considerations}, we discuss whether a more informative prior on $f_w$ could partly solve this issue.

	The resulting $\eta_\mathrm{2ph}$ for the two-phase gas range between $\approx 2 \times 10^{-4}$ and $0.071$ with a median of $\langle \eta_\mathrm{2ph}\rangle \approx 0.012$ (see Table~\ref{tab:supernova_efficiencies}).
	The fraction of WNM also varies greatly between galaxies, as expected given the large variations of $\eta$. We obtain WNM fractions between $f_\mathrm{w} \approx 0.16$ and $f_\mathrm{w} \approx 0.97$ with a median of $\langle f_w\rangle \approx 0.41$. 
	We conclude that, if both the SN efficiency and the WNM fraction are allowed to vary across galaxies, there is no obvious preference for HI being dominated by the WNM. 
	This is somehow in contradiction with previous studies, which concluded that the WNM fraction is large in most galaxies and is expected to be even higher in dwarfs due to their metal-poor and diffuse ISM (e.g. \citealt{2003_Heiles_Troland_WNM,2013_pineda,Young_2003_WNM_dIrrs,Dickey_2000,2000-zimmer}). 
	The right panel in Fig.~\ref{fig:All_IN} shows the maximum SN energy versus turbulence energy for the two-phase case. 
	Compared to the 100\% WNM case, the scatter is reduced, and indeed the majority of the data points are within two orders of magnitude in efficiency ($0.001 \lesssim \eta \lesssim 0.1$). 
	This suggests that the two-phase case is more suitable to describe the observations. 
	We note that the outliers are galaxies showing a strong degeneracy between $\eta$ and $f_w$. 
	
	In summary, although the characteristic SN efficiencies for each dwarf span a wide range across the three ISM cases, the required efficiencies are always well within the limit of 10\% predicted by SN remnant models, indicating that SN feedback is a viable driver of HI turbulence in dwarfs. The galaxy-to-galaxy variation in $\eta$, however, raises the question of whether a single efficiency can simultaneously explain the turbulence across the full sample. We address this in the following section. 
	
	\subsection{"Universal" SN efficiency}\label{sec:GlobalEta_intro}
	Finally, we test the hypothesis of a "universal" SN efficiency across our sample of galaxies. 
	This is done by imposing that $\eta$ is the same for all galaxies when we explore the parameter space. 
	We perform a fit of all galaxies together to determine this "global" efficiency ($\eta_\mathrm{gb,2ph}$), along with the values of $f_w$ for each individual galaxy. This choice is motivated by the fact, while the fraction of WNM could in principle vary between galaxies depending on their properties (e.g. local SFR, metallicity), the SN efficiency may not, as it mostly depends on the energy radiated away during the SN remnant evolution \citep{2019CFN}. 
	
	\subsubsection{Considering dwarf and spiral samples separately}\label{sec:GlobalEta_separated}
	We first analyse our samples of dwarfs and spirals separately, exploring the three cases with different $f_\mathrm{w}$.  
	The results of these explorations are reported in Table~\ref{tab:supernova_efficiencies}. 
	Unsurprisingly, the SN efficiency decreases with increasing contribution of the WNM. 
	For comparison, Table~\ref{tab:supernova_efficiencies} also reports the efficiency for the spirals from \citetalias{2020_bacchini_SN_feedback} only, re-analysed using the non-centred approach described in Sect.~\ref{sect:methods}. 
	The efficiencies for spirals are consistent within the uncertainties with the median values reported by \citetalias{2020_bacchini_SN_feedback}. 
	Overall, the SN efficiencies for dwarfs only are lower than those for spirals only, with values consistent within $3 \sigma$ only for the WNM and the two-phase cases. This suggests that it is possible to find a single "universal" efficiency when considering the two samples together. Such hypothesis is explored in Sect.~\ref{sec:GlobalEta}.
	Figure~\ref{fig:EnergyComponents-2phase-global} shows the radial profiles of the energy contributions for dwarfs only in the case of a two-phase HI, which  gives efficiency of $\approx 0.013$. 
	The energy profiles are reproduced remarkably well, and the same is for the two extreme cases not displayed here. 
	\begin{table}[ht]
		\centering
		\caption{"Universal" supernova efficiencies for different galaxy samples and ISM assumptions.}
		\begin{tabular}{lccc}
			\hline
			\hline
			& Spirals & Dwarfs & All \\
			\hline
			$\eta_{\text{gb,CNM}}$ & $0.033^{+0.002}_{-0.002}$ & $0.026^{+0.001}_{-0.001}$ & $0.030^{+0.001}_{-0.001}$ \\
			$\eta_{\text{gb,WNM}}$ & $0.014^{+0.001}_{-0.001}$ & $0.001^{+0.001}_{-0.001}$ & $0.008^{+0.001}_{-0.001}$ \\
			$\eta_{\text{gb,2ph}}$ & $0.019^{+0.002}_{-0.001}$ & $0.013^{+0.002}_{-0.002}$ & $0.017^{+0.001}_{-0.001}$ \\
			\hline
			\hline
		\end{tabular}
		\label{tab:supernova_efficiencies}
	\end{table}
	
	When compared to the characteristic SN efficiencies for the two-phase case (Fig.~\ref{fig:Energycomponents-2phase-single}), the "universal" SN efficiency typically differs only by a factor of 2, with the WNM fraction increasing or decreasing to balance the reduced or enhanced contribution from SNe. The most drastic change occurs in DDO~168, which clearly illustrates the degeneracy problem. In DDO~168, the SN efficiency found in this case is $\eta_\mathrm{gb,2ph} \approx 0.013$, much lower than the value $\eta_\mathrm{2ph} \approx 0.071$ found in Sect.~\ref{sec:2phase_single}. 
	This is because the WNM fraction increases from $f_\mathrm{w} \approx 0.180$ to $f_\mathrm{w} \approx 0.907$ to compensate for the different contribution of SN energy in the two cases. 
	In contrast to the characteristic SN efficiency, the resulting posteriors for both $\eta_\mathrm{gb,2ph}$ and $f_w$ are very well constrained in all three scenarios. This is because a larger amount of data is available to infer the free parameters of interest. This effect is therefore primarily seen in dwarfs, where some galaxies contain a small number of data points, though a few spiral galaxies also exhibit improved posteriors.
	
	\subsubsection{Considering dwarfs and spirals together}\label{sec:GlobalEta}
	Lastly, we explore the case of a single, "universal" efficiency, the same for both dwarfs and spirals. 
	We recall that, when considered together, our samples are representative of the main sequence of nearby star-forming galaxies. 
	As done in Sect.~\ref{sec:GlobalEta_separated}, we explore the CNM, WNM, and two-phase case. 
	We report the resulting efficiencies in Table~\ref{tab:supernova_efficiencies}. 
	
	Figure~\ref{fig:GlobalEfficiency_2phase} shows the radial profiles of the observed turbulent energies versus the maximum SN energy, for both spirals and dwarfs. 
	As expected, when we impose a "universal" SN efficiency, the data points for dwarfs and spirals are less scattered than in Fig~\ref{fig:All_IN} for the characteristic efficiencies. 
	We also note that the data points that seem to require SN efficiency much higher than the "universal" value found here (i.e. $0.017$) belong to the disc outskirts and, when their uncertainties (not shown) are considered, they are still consistent with the "universal" efficiency. 
	In Appendix~\ref{app:radial_correlation}, we show that this result is not driven by similar radial trends of the SN and turbulent energies. 
	
	We conclude that hypothesis of "universal" SN efficiency is fully reasonable and can explain the HI turbulent energy across the disc of star-forming galaxies along the main sequence and, we recall, even for dwarf starbursts. 
	Based on our experiment for a two-phase medium, we find  that only $\approx 0.02$ (i.e. $\approx 2\%$) of SN energy is needed to drive the HI turbulence across the star-forming sequence and that this value is within a factor 2 against the uncertainties due to the actual abundances of CNM and WNM in the HI. 
	This value is remarkably similar to that predicted by models of SN remnant evolution, which is indeed $\approx 0.02$ (e.g. \cite{2019CFN}, Sect.~8.7). 
	Our results are statistically more robust, yet in agreement, with the median values of the characteristic SN efficiencies obtained by \citetalias{2020_bacchini_SN_feedback}.  
	
	\section{Discussion}\label{sect:discussion}
	In this section, we discuss the results and possible caveats of this work. We anticipate that caveats do not impact our main conclusion: even in regions with low $\Sigma_\mathrm{SFR}$ (e.g. dwarfs), the observed atomic gas turbulence can be maintained by SN feedback alone with injection efficiency $\eta$ between $4\times10^{-5}$ and $0.087$.
	
	\subsection{Further implications}\label{sec:self-regulating-cycle}
	Our findings fit into a broader theoretical framework in which SN feedback is a key player in the self-regulating cycle of star formation in the ISM (e.g. \citet{Ostriker_2011}, \cite{2022_ostriker_kim}). 
	In this cycle, stellar feedback, especially via SNe, injects energy into the ISM, increasing the gas turbulence and pressure. 
	The gas disc thickness is set by this feedback-driven pressure, which counteracts gravity in a quasi-steady balance. When gravity prevails, the gas disc is thin and dense, and has high SFR (per unit volume). 
	When massive stars start exploding as SNe, feedback increases pressure and turbulence, so the gas disc becomes thicker and less dense, and thereafter the SFR decreases. As turbulence fades due to dissipation, gravity dominates again over pressure, the disc gets thinner, and the cycle restarts with a new phase of high SFR.
	
	The strong correlation between the SFR and gas densities per unit volume is found in \citetalias{Bacchini_2019a}, \citetalias{2019b_bacchini}, and \citetalias{Bacchini_2020_volumetric}, respectively, for spirals and dwarfs using essentially the same samples of galaxies used in this work. 
	This correlation, called the volumetric star formation law, was obtained by converting the observed surface densities into volume densities using the scale height of the gas disc and assuming the balance between the gas pressure, given by the observed velocity dispersion of the gas, and gravity. 
	Our work and \citetalias{2020_bacchini_SN_feedback} complement these empirical results by demonstrating that the gas velocity dispersion, which regulates the gas disc thickness through the pressure-gravity balance, is consistent with being driven by SN feedback only, in both dwarf and spiral galaxies. 
	We also note that the SN efficiency found in this work (as well as in \citetalias{2020_bacchini_SN_feedback}) implies that only $\approx 2\%$ of the energy from stellar feedback is spent to drive the HI turbulence.
	
	It is instructive to compare this value with the fraction of $E_\mathrm{SN}$ that models and simulations predict to survive as kinetic energy of the surrounding gas, a natural proxy for the energy available to drive turbulence: for a remnant evolving in a radiatively cooling medium this is $\approx 10\%$ \citep[e.g.][]{2015_SN_energy_kim_ostriker,Fierlinger_SNR,LauBonnell2025}, i.e. a factor $\approx 5$ larger than what our data require.
	This leaves plenty of energy to drive other galactic phenomena ascribed to SNe and, more in general, to stellar feedback, such as ionised gas winds in dwarfs (e.g. \citealt{2023Marasco}) or galactic fountains observed in spirals \citep{2019_Marasco}. 
	
	\begin{figure}[ht]
		\centering   \includegraphics[width=1\linewidth]{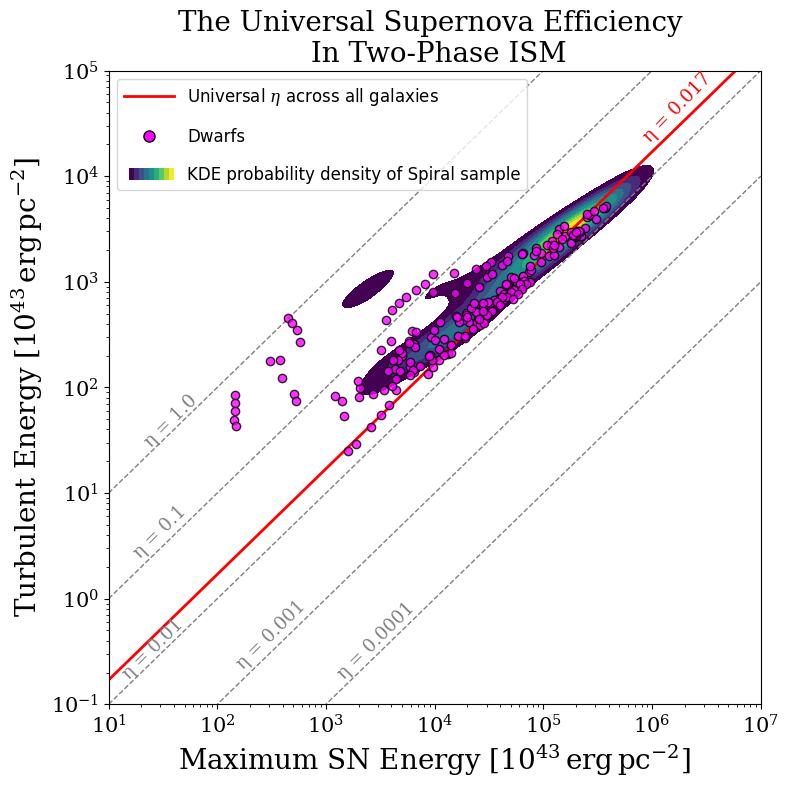}
		\caption{Same as Fig.~\ref{fig:All_IN}, but for the case of a "universal" SN efficiency for both dwarfs (points) and spirals (contours), as indicated by the solid red line. Both axes show energies per unit area.} 
		\label{fig:GlobalEfficiency_2phase}
	\end{figure}

	\subsection{Comparison with previous works}\label{sec:comparison}
	A few previous studies have investigated the origin of the HI turbulence in galaxies, considering both the role of stellar feedback and other possible mechanisms. 
	
	\citet{Stilp_2013_drivers_of_turbulence} analyse $\sigma_\mathrm{HI}$ measured across multiple dwarf and low-mass spiral galaxies using "super-profiles", which are created by stacking the line profiles from different pixels across each galaxy after removing the rotational velocity contribution. They find a correlation between $\Sigma_\mathrm{SFR}$ and the HI turbulence in the regions where $\Sigma_\mathrm{SFR} \gtrsim 10^{-3} \ \mathrm{M}_\odot \ \mathrm{yr}^{-1} \ \mathrm{kpc}^{-2}$, but this correlation weakens where $\Sigma_\mathrm{SFR}$ is lower. 
	In this low $\Sigma_\text{SFR}$ regime, SN feedback would require unrealistically high efficiencies ($\eta \geq 1$), so they conclude that SNe likely contribute, and may even set a lower bound for $\sigma_\mathrm{HI}$, but that some additional mechanism must operate. They consider other candidates (e.g. MRI), but none could fully account for the turbulence in dwarfs, mainly because their shallow potential wells and solid-body rotation curves limit the formation of shear.
	
	There are at least two fundamental differences between our work and \citet{Stilp_2013_drivers_of_turbulence} that can explain why we reach completely different results. The first is that this previous work neglected the effects of the HI flaring, which prolongs the turbulence dissipation time, thus reducing the energy supply to sustain the HI turbulence exactly in the regions where $\Sigma_\text{SFR}$ is low. \citet{Stilp_2013_drivers_of_turbulence} assume a constant scale height across all radii, which gives a uniform dissipation timescale 45~Myr. This is less than half the dissipation timescales adopted here (Fig.~\ref{fig:DissipationTime}), meaning that more energy must be injected per unit time to maintain a stable turbulence state. 
	We also highlight the recent study, \citet{2025_hunter_timescales}, which supports the idea of dissipation timescales of $\gtrsim 100$~Myr. 
	They analyse 26 low-mass galaxies with resolved stellar populations and partially overlapping with our sample. 
	They find significant correlation between the HI turbulence measures and the star formation activity
	$\gtrsim 100$~Myr ago, suggesting the star formation activity and HI are coupled on such timescales. They argue that this may be related to the time scales over which turbulence decays in the ISM.
	The second difference is how $\sigma_\mathrm{HI}$ is measured. As anticipated in Sect.~\ref{sect:kinematics}, stacking line profiles can add artificial broadening and exacerbate beam smearing, inflating $\sigma_\mathrm{HI}$ and thus biasing the SN efficiencies high, whereas our measurements are corrected for beam smearing.
	
	\citet{Hunter_2021_correlation} also investigate the impact of stellar feedback on HI in dwarfs, leveraging essentially the same data we have used in this work and for a sample of galaxies overlapping with ours. They perform a 2D (i.e. pixel-by-pixel) study of the spatial correlation between star formation and HI turbulence, using maps of the FUV emission and of the HI kinetic energy density (KED) and $\sigma_\mathrm{HI}$.
	These authors measure, on spatial scales of $\approx 20-340$~pc depending on galaxy distance, how much excess KED appears locally around star-forming regions above the radial average of a galaxy. 
	They find that the SN efficiency needed locally, which they argue is the most direct measurement possible, is $\approx 2$ times too small to explain the turbulence of the whole galaxy, based on the efficiencies measured by \citetalias{2020_bacchini_SN_feedback} for spirals, and conclude that other mechanisms, possibly of gravitational origin, must contribute. As complementary evidence, they note that the $\sigma_\mathrm{HI}$ excess around star-forming regions is only $\approx 0.37$~km/s, tiny compared to the typical $\sigma_\mathrm{HI} \approx 6-10$~km/s and the associated uncertainties. 
	
	The tension between our work and \citet{Hunter_2021_correlation} has two possible explanations, one methodological and one physical. Methodologically, we and \citetalias{2020_bacchini_SN_feedback} work exclusively with radial profiles: azimuthal averaging increases the S/N, which is crucial to robustly measure $\sigma_\mathrm{HI}$ and to account for the HI disc flaring, but it smooths the local variations. 
	Measuring $\sigma_\mathrm{HI}$ pixel by pixel, however, suffers from beam smearing, low S/N, or local features such as winds and outflows. 
	Lastly, \citet{Hunter_2021_correlation} cannot cleanly separate the thermal contribution, and so apply approximate correction factors; and they adopt an averaged HI scale height, although this quantity varies both between and within galaxies. We instead treat the thermal contribution in a statistically robust way and use the radial profile of the scale height.
	Nonetheless, the tension with \citet{Hunter_2021_correlation} may not be a genuine contradiction, as we probe fundamentally different physical scales. 
	They test whether turbulence is enhanced locally and instantaneously at star-forming sites, while our approach can probe the SN efficiency needed across the whole disc and, because of the azimuthal averaging, in a more time-averaged sense. 
	
	We also stress that, even if stellar feedback were insufficient, gravitational instabilities are unlikely to be a viable alternative on scales of a few 100~pc: \citetalias{bacchini20243dviewlocalgravitational} find our dwarfs to be far from gravitationally unstable, thanks to the stabilising effect of the HI disc thickness and of the DM halo. We cannot exclude a role on $\sim 10$~pc scales, but testing this would require 21-cm data of much higher S/N and resolution than used here.
	
	\subsection{Stability of results}
	We now discuss the stability of our results against the underlying assumptions. 
	A discussion about the impact of the IMF choice, the assumption of $L_D = 2 h$ or $L_D = h$, and other potential sources of turbulence can be found in \citetalias{2020_bacchini_SN_feedback}. Here, we limit the discussion to aspects that are more important for dwarfs than spirals.  
	
	\subsubsection{Mass models}
	With the approach used here, $h_\mathrm{HI}$ is mostly sensitive to changes in the velocity dispersion, while changes in the gravitational potential have a minor effect or give consistent scale height within the uncertainties (see Appendices~A and~E in \citetalias{Bacchini_2019a}). 
	\cite{2022ManceraPina_b} have explored the effects of the gas disc flaring on the recovered dark matter halo parameters from the HI rotation curve decomposition, finding minor effects for most galaxies analysed here. The only notable exception is the very gas-dominated galaxy CVn~I~dwA, for which they find $\approx 2$ times lower DM halo mass than the one used. This would reduce $h_\mathrm{HI}$ by $\lesssim 30$\%, thus facilitate the sustaining of turbulence. 
	Nevertheless, such small changes have negligible effects on the final SN efficiency, as the uncertainties on observables are larger. 
	The same argument applies for the changes in the DM halo parameters that arise from improving the constraints on the stellar distribution and assuming more flexible priors (for a thorough study, see \citealt{2025ManceraPina}). 
	
	\subsubsection{Mass limits for SN explosions}
	The value of $f_\mathrm{cc}$ depends on the upper mass limit of the IMF, which we set to 120 $M_\odot$. Stars significantly exceeding this value do exist \citep[e.g.][]{crowther, yusof2010lifedeathmassivestars}, but they are extremely rare and lie on the very steep tail of the IMF, so their impact is negligible. Conversely, our limit might be too high, since stars above a certain threshold are expected to collapse directly to black holes rather than explode as SNe; \cite{2025_maltsev} suggest $\approx 30 \ {M_\odot}$, which would only decrease $f_\mathrm{cc}$ from $1.3 \times 10^{-2} M^{-1}_\odot$ to $1.1 \times 10^{-2} M^{-1}_\odot$. The choice of upper mass limit is therefore not critical. 
	In future studies these uncertainties could be included through appropriate hyperpriors on $f_\mathrm{cc}$.
	
	\subsubsection{Number of data points per galaxy}
	Spirals have larger HI discs than dwarfs, hence more resolution elements and more data points. 
	Spirals might thus weigh more than dwarfs in the determination of the "universal" SN efficiency and bias it towards the spiral value. This could be mitigated by weighting the data by the number of resolution elements, or by interpolating the profiles to impose the same number of points per galaxy, but the bias is in any case weakened by our sample containing fewer spirals (nine) than dwarfs (14). We tested this directly by re-running the algorithm on sub-samples with several spirals (or dwarfs) removed: the inferred $\eta$ and $f_\mathrm{w}$ remain essentially unchanged.
	
	\subsubsection{Choice of upper limits on the SFR surface density}
	In order to have a conservative upper limit on $\Sigma_\mathrm{SFR}$ in regions where FUV emission is not detected, we extrapolate the $\Sigma_\mathrm{SFR}$ up to the largest radius with measured HI distribution and kinematics assuming an exponential profile (Sect.~\ref{sec:sample}). 
	This extrapolation introduces an uncertainty in the outermost regions of the disc, where $\tau_\mathrm{d}$ is longest and the WNM contribution is most important. Moreover, this approach is not fully consistent with that used to set upper limits for spirals, which are obtained by converting the detection limit reported by \cite{Leroy_2008}. 
	To assess the sensitivity of our results to this choice, we repeated the analysis excluding all radial bins where $\Sigma_\mathrm{SFR}$ is an upper limit rather than a direct measurement. The inferred "universal" SN efficiencies change by less than $\approx 0.001$, suggesting a minor impact of lower limits on our results. We also note that our choice is conservative because, if the upper limits are underestimated, the SN efficiencies would be overestimated, reinforcing our conclusion that SNe alone are sufficient for driving turbulence.
	
	\subsubsection{Distribution and composition of the ISM}
	Our method assumes that HI is distributed in two phases, WNM and CNM, and that it is thermally stable. However, observations suggest part of the atomic gas may be thermally unstable, e.g. \citet{2003_Heiles_Troland_WNM}, \citet{Roy_2013}, \citet{Kalberla_2018}. Even though this would not influence the resulting SN efficiency, it implies that our estimate of $f_\mathrm{w}$ is an upper limit, as part of the thermal energy could be partially ascribed to thermally unstable HI at intermediate temperatures. 
	Furthermore, we assume that $f_\mathrm{w}$ is constant across the disc. 
	This choice may not be fully realistic, especially in dwarfs, where the radiative heating is promoted by the low metallicity and density of the ISM. 
	As discussed in \citetalias{2020_bacchini_SN_feedback}, modelling the radial gradient  $f_\mathrm{w}$ is not advisable, as it would introduce additional free parameters and worsen the degeneracies. 
	To solve these issues, one may attempt to specifically model $f_\mathrm{w}(R)$ and include additional observations to constrain the model parameters. 
	However, this is beyond the scope of this work and is left to future studies. 
	
	The observed velocity dispersions are measured on spatial scales of $\sim 200 \ \mathrm{pc}$, implying that each line profile represents a composite of velocities originating from the CNM, WNM, and large-scale turbulent motions along the line of sight. 
	Since the vertical gas distributions are derived from these observed dispersions without decomposing the CNM and WNM contributions, we effectively assign a common pressure support, and hence a shared scale height, to both phases. Assuming thermal equilibrium, we thus treat the neutral ISM as a perfect mix of two phases with different temperature and turbulence, which, as noted in Sect.~\ref{sec:mass_models}, may be an oversimplification if the two phases do not share the same velocity dispersion and vertical distribution. 
	If we leave behind the assumption of thermal equilibrium, the $f_w$ values found in the two-phase case can be interpreted in terms of thermal velocity of a single-phase neutral gas with intermediate temperature. A fully thermally unstable HI is, however, unrealistic, as observations suggest that a fraction of HI might be unstable \citep{2003_Heiles_Troland_WNM,2018Murray}.
	
	\subsubsection{The environment of SN explosions}\label{sect:sn_environment}
	Our model does not assume the density of the medium in which individual SNe explode: $\eta$ is a free parameter, fitted against the HI distribution and kinematics observed at each radius. Our assumption is rather that the azimuthally averaged gas and SFR surface densities are representative of the mean conditions at that radius, over the timescales relevant for star formation and feedback. Our analysis therefore concerns the azimuthally averaged energy balance in radial annuli rather than the surroundings of individual SNe: $\eta$ is an effective efficiency averaged over the explosion sites contributing to a given annulus, and local environmental inhomogeneities are expected to average out over these scales. The idealisation lies not in the ambient density itself, but in describing with a single $\eta$ a population of remnants evolving in different environments. Since the energy retained by a remnant might depend on the local density \citep[e.g.][]{LauBonnell2025}, and massive stars typically form in clusters \citep{Facchini2026}, one may ask whether this effective mean is biased. 
	
	Using evolved massive stars in M33 as proxies for explosion sites, \citet{Sarbadhicary2026} find a mass-dependent correlation between progenitors and dense gas, so a fraction of core-collapse SNe is expected to explode where remnants are confined by dense gas, cool early and deliver little kinetic energy \citep[e.g.][]{LauBonnell2025}. 
		However, \cite{Sarbadhicary2026} also find that at least $45\%$ of Wolf--Rayet stars and up to $77\%$ of red supergiants have no detectable $\mathrm{H_2}$ at their location, while more than $90\%$ are associated with detectable HI. This supports a picture in which winds, ionising radiation, and pre-SN feedback clear much of the parent cloud before many core-collapse SNe explode, so that a substantial fraction of SNe injects its energy into gas far less dense than the clouds where the progenitors formed. 
	
	Consistently, \citet{LauBonnell2025} find that their "semi-confined" case, arguably the most realistic since pre-SN feedback carves "channels" out of the parent cloud, yields an energy budget similar to the free-field case, whereas only the fully "confined" case is strongly suppressed. Such channels also let the energy escape the dense star-forming regions, consistent with the small local excess of $\sigma_\mathrm{HI}$ reported by \citet{Hunter_2021_correlation} (Sect.~\ref{sec:comparison}). 
		A closer analogue of the fully "confined" case, in which the remnant evolves inside dense molecular gas, could be the molecular phase itself. 
		This was analysed by \citetalias{2020_bacchini_SN_feedback}, who derive the efficiency needed to sustain the turbulence of the molecular gas in the same sample of spirals studied here, finding efficiencies of the order of $0.1\%$, much closer to the yield from confined SNe in \cite{LauBonnell2025} than for HI. 
	
	Our conclusions therefore do not rely on SNe exploding predominantly in low-density gas: even in the densest phase, the required efficiency stays well below the limit set by SN remnant models. Finally, recovering a similar $\eta$ across a sample spanning dwarfs (including a couple of starbursts) and spirals is itself an empirical indication that this effective efficiency varies little between galaxies.

	\subsubsection{Energy losses due to shocks}\label{sect:shocks}
	Our approach relies on the assumption that the turbulent energy is conserved from the injection scale down to the physical scales probed by the HI observations (i.e. the scale at which $\sigma_\mathrm{HI}$ is measured, corresponding to a few 100~pc \citet{Iorio_2017_data,Bacchini_2019a}). 
	This assumption of energy conservation is at the basis of the \cite{1941Kolmogorov} theory for turbulence, which assumes an incompressible fluid. 
	In this framework, the turbulent flow is made up of eddies covering a wide range of spatial scales. The turbulent energy is initially injected at the driving scale ($L_D$), the size of the largest eddies. These then fragment into progressively smaller eddies, thereby transferring energy to smaller scales. Throughout this cascade, the energy is conserved until the dissipation scale is reached. Then, viscous effects convert the turbulent (kinetic) energy into internal energy. 
	
	In a more realistic picture, the ISM should be considered compressible, and supersonic motions are expected to cause shocks, leading to radiative losses in the turbulent cascade. In this eventuality, the SN efficiency obtained in this work should be interpreted as a lower limit. 
	We argue that radiative losses would not change our conclusions dramatically based on three considerations. 
	First, compressible motions seem to constitute 30\%-50\% of the ISM turbulence, meaning our efficiencies are robust within a factor 2-3 \cite[][]{2016_padoan,2016_Pan,2017_Orkisz}. 
	Second, the power spectrum of HI intensity fluctuations in dwarfs shows slopes between $\approx -1.5 \ \text{and} -2.6$ \citep{Dutta_2009}, while V-band and H$\alpha$ emission over scales of 10 - 400 pc show slopes of $\approx -1.4 \pm 0.3$ \citep{2005_willet_dwarf_vs_kolmo}. These values are comparable to the $k^{-5/3}$ power law scaling expected for the Kolmogorov turbulence, which might suggest that HI can be partially approximated as incompressible. 
	Third,  when we consider the most realistic scenario of a two-phase HI, we typically find that the HI turbulence is at most slightly supersonic, with Mach number $\mathcal{M} = v_\mathrm{turb} /v_\mathrm{th}$ in the range $1 \lesssim \mathcal{M}\lesssim 2$ for both spirals and dwarfs. 
	When we explore the "universal" SN efficiency, some galaxies (e.g. DDO~50, DDO ~87) do show supersonic motions ($\mathcal{M}\approx 2 - 6$), but mostly in the inner regions of the galactic disc.
	This suggests that radiative losses might be secondary at these spatial scales, although the large uncertainties on the effective distribution of HI temperature (and thus on $\mathcal{M}$) prevent us from ruling out radiative shocks.
	
	\subsubsection{Other considerations}\label{sect:considerations}
	In a hierarchical Bayesian model, the priors should be informative enough to guide the sampling while leaving the model some flexibility. Ours follow \citetalias{2020_bacchini_SN_feedback} (Sect. \ref{sect:methods}): the prior on $\eta$ is a beta-function giving equal weight to low and high values and less weight to intermediate ones, motivated by previous studies where $\eta$ tended to be either very high or very low depending on the galactic environment.
	The prior on the second free parameter of interest, $f_w$ was derived from
	\begin{gather}
		v_\mathrm{th} = \sqrt{(1-f_w)v_{\mathrm{th,c}}^2 + f_wv_{\mathrm{th,w}}^2} \\
		\Rightarrow f_w = \frac{v_\mathrm{th}^2 - v_\mathrm{th,c}^2}{v_\mathrm{th,w}^2 - v_\mathrm{th,c}^2} = \frac{\mathrm{Uniform}[1 \mathrm{km/s},8.1 \mathrm{km/s}]^2 - (1 \mathrm{km /s})^2}{(8.1\mathrm{km/s})^2 - (1\mathrm{km/s})^2}\label{eq:fw2}
	\end{gather}
	This choice of prior for $f_w$ is thus skewed towards smaller values due to the nonlinear (quadratic) dependence of Eq.~\ref{eq:fw2}. Nonetheless, this does not influence the main conclusions of this work, since the thermal energy would become relatively more important while lowering the required SN efficiency to drive the observed turbulence. We also consider the two limiting ISM cases independent of the $f_w$ parameter. A more informative choice of prior could be beneficial for more efficient exploration of the parameter space. More in-depth studies of the fraction of WNM in dwarfs are required before we can update the chosen priors.
	The main outcomes of this study are therefore only weakly dependent on the specific priors and on the adopted methodology. We also cross-validated the Bayesian results against the analytical solution for the radial profile of $\eta$, assuming $f_w = 0, 50, 100 \%$: in all three cases $\eta$ typically stays below $0.1$, and where this limit is occasionally exceeded in the outer disc the values remain consistent within the uncertainties.
	
	\section{Conclusion}\label{sect:conclusion}
	This study investigates the role of SN feedback in maintaining HI turbulence in dwarf irregular galaxies, and the existence of a "universal" SN efficiency across the star-forming main sequence. From the observed HI distribution and kinematics, we derived the HI kinetic energy per unit area across their discs, and reproduced these profiles with a model in which HI is in hydrostatic equilibrium and its kinetic energy is the sum of the turbulent energy from SNe, traced by the SFR distribution from UV images, and the HI thermal energy.
	This is done considering that the flaring of the HI disc prolongs the dissipation timescale of turbulence and reduces the energy required to sustain turbulence, especially towards the disc outskirts \citepalias{2020_bacchini_SN_feedback}. This is a key difference from previous works investigating the origin of HI turbulence in dwarfs (e.g. \citealt{Stilp_2013_drivers_of_turbulence}, \citealt{Hunter_2021}), which often neglect that the HI scale height varies both within and across galaxies. 
	We use a hierarchical Bayesian inference to measure the SN efficiency $\eta$, the fraction of SN energy needed to sustain the observed HI turbulence. 
	The key findings of this paper are:
	\begin{itemize}
		\item The HI turbulence in nearby dwarf galaxies is consistent with being driven by the SN feedback only, with no mandatory contribution from other sources of turbulence.  In the most realistic scenario of two-phase HI, we find $\eta_\mathrm{gb,2ph} \approx 0.013$. Even when the HI thermal energy is assumed negligible (CNM case), we find SN efficiencies below $\approx 9$\% when each dwarf is considered singularly, and $\eta_\mathrm{gb,CNM} \approx 3\%$ when we impose that dwarfs must have the same SN efficiency. 
		
		\item The HI thermal motions significantly contribute to the observed kinetic energy, especially in the outskirts of galaxies where the HI is diffuse and $\Sigma_\mathrm{SFR}$ is low. However, it remains a challenge to disentangle the thermal motions from the turbulence. Although we can constrain $f_w$ well by imposing a global efficiency $\eta_\mathrm{gb}$, it is clear that the degeneracy between $\eta$ and $f_w$ remains unbroken for some galaxies.
		
		\item A single "universal" SN efficiency of $\eta_\mathrm{gb,2ph} \approx 0.02$ ($\approx 2$\%) can simultaneously reproduce the observed HI kinetic energy profiles for our sample of dwarfs and spirals, which is representative of the present-day star-forming main sequence. 
	\end{itemize}
	We estimate a systematic uncertainty on the "universal" efficiencies of a factor $\approx 2$, primarily due to the uncertain HI thermal state. This does not change our conclusion that the fraction of SN energy required to sustain the HI turbulence is within physically motivated limits of $\approx 10\%$. 
	A SN efficiency of $\approx 2$\% is remarkably similar to that expected from models of SN remnant evolution \citep[e.g.][Sect.~8.7]{2019CFN}. 
	Our study offers a useful benchmark for numerical and semi-analytical models of galaxy formation and evolution, which must rely on sub-grid recipes for stellar feedback since single SNe cannot be resolved. 
	The "universal" SN efficiency obtained here could serve as an empirically-motivated value for setting up the fraction of SN energy that is transferred to the cold ISM in these sub-grid models, reducing the degrees of freedom. 
	In the future, this study could be extended in two directions. One is the high redshift Universe, to test whether the SN efficiency is truly universal and remains unvaried across cosmic time. This is crucial to understand and model galaxy evolution. Second, the Square Kilometre Array may allow to push the studies on the origin of turbulence to even more extreme environments, such as gas-rich ultra-diffuse galaxies and the edges of low-mass galaxies. 
	
	\section*{Data availability}
	The atlas of the posterior distributions is available on Zenodo at \url{https://doi.org/10.5281/zenodo.23214489}.

	\begin{acknowledgements}
		The authors thank P.~E.~Mancera-Pi\~na for useful comments that helped improving this work. 
		LS acknowledges the support from the Erasmus program 2024-2025 and the hospitality of the DARK section of the Niels Bohr Institute, University of Copenhagen. CB acknowledges support from the Carlsberg Foundation Fellowship Programme by Carlsbergfondet.
	\end{acknowledgements}

	\bibliography{library}

@BOOK{2009_stellar_ev,
       author = {{Prialnik}, Dina},
        title = "{An Introduction to the Theory of Stellar Structure and Evolution}",
    publisher = {Cambridge University Press},
         year = {2009},
       adsurl = {https://ui.adsabs.harvard.edu/abs/2009itss.book.....P}
}

@article{2020_bacchini_SN_feedback,
	author = {{Bacchini}, C. and {Fraternali}, F. and {Iorio}, G. and {Pezzulli}, G. and {Marasco}, A. and {Nipoti}, C.},
	title = {Evidence for supernova feedback sustaining gas turbulence in nearby star-forming galaxies},
	DOI= "10.1051/0004-6361/202038223",
	url= "https://doi.org/10.1051/0004-6361/202038223",
	journal = {\aap},
	year = {2020},
	volume = {641},
	pages = "A70",
}

@article{Bacchini_2020_volumetric,
   title={The volumetric star formation law for nearby galaxies: Extension to dwarf galaxies and low-density regions},
   volume={644},
   ISSN={1432-0746},
   url={http://dx.doi.org/10.1051/0004-6361/202038962},
   DOI={10.1051/0004-6361/202038962},
   journal={\aap},
   publisher={EDP Sciences},
   author={Bacchini, Cecilia and Fraternali, Filippo and Pezzulli, Gabriele and Marasco, Antonino},
   year={2020},
   month={dec},
   pages={A125}}

@ARTICLE{bacchini20243dviewlocalgravitational,
       author = {{Bacchini}, C. and {Nipoti}, C. and {Iorio}, G. and {Roman-Oliveira}, F. and {Rizzo}, F. and {Mancera Pi{\~n}a}, P.~E. and {Marasco}, A. and {Zanella}, A. and {Lelli}, F.},
        title = "{A 3D view on the local gravitational instability of cold gas discs in star-forming galaxies at 0 {\ensuremath{\lesssim}} z {\ensuremath{\lesssim}} 5}",
      journal = {\aap},
         year = 2024,
        month = jul,
       volume = {687},
          eid = {A115},
        pages = {A115},
          doi = {10.1051/0004-6361/202449925},
archivePrefix = {arXiv},
       eprint = {2405.00103},
 primaryClass = {astro-ph.GA},
       adsurl = {https://ui.adsabs.harvard.edu/abs/2024A&A...687A.115B}
}

@ARTICLE{2019b_bacchini,
       author = {{Bacchini}, C. and {Fraternali}, F. and {Pezzulli}, G. and {Marasco}, A. and {Iorio}, G. and {Nipoti}, C.},
        title = "{The volumetric star formation law in the Milky Way}",
      journal = {\aap},
         year = 2019,
        month = dec,
       volume = {632},
          eid = {A127},
        pages = {A127},
          doi = {10.1051/0004-6361/201936559},
archivePrefix = {arXiv},
       eprint = {1911.03480},
 primaryClass = {astro-ph.GA},
       adsurl = {https://ui.adsabs.harvard.edu/abs/2019A&A...632A.127B}
}

@ARTICLE{Mogotsi_2016,
       author = {{Mogotsi}, K.~M. and {de Blok}, W.~J.~G. and {Cald{\'u}-Primo}, A. and {Walter}, F. and {Ianjamasimanana}, R. and {Leroy}, A.~K.},
        title = "{H I and CO Velocity Dispersions in Nearby Galaxies}",
      journal = {\aj},
         year = {2016},
        month = jan,
       volume = {151},
       number = {1},
          eid = {15},
        pages = {15},
          doi = {10.3847/0004-6256/151/1/15},
archivePrefix = {arXiv},
       eprint = {1511.06006},
 primaryClass = {astro-ph.GA},
       adsurl = {https://ui.adsabs.harvard.edu/abs/2016AJ....151...15M}
}

@article{zhang2012_stellarmass,
  title={OUTSIDE--IN SHRINKING OF THE STAR-FORMING DISK OF DWARF IRREGULAR GALAXIES},
  author={Zhang, Hong-Xin and Hunter, Deidre A and Elmegreen, Bruce G and Gao, Yu and Schruba, Andreas},
  journal={\aj},
  volume={143},
  number={2},
  pages={47},
  year={2012},
  publisher={IOP Publishing}
}

@ARTICLE{1931_chandrasekhar,
       author = {{Chandrasekhar}, S.},
        title = "{The Maximum Mass of Ideal White Dwarfs}",
      journal = {\apj},
         year = {1931},
        month = jul,
       volume = {74},
        pages = {81},
          doi = {10.1086/143324},
       adsurl = {https://ui.adsabs.harvard.edu/abs/1931ApJ....74...81C}
}

@article{Hunter_2010_data,
   title={GALEXULTRAVIOLET IMAGING OF DWARF GALAXIES AND STAR FORMATION RATES},
   volume={139},
   ISSN={1538-3881},
   url={http://dx.doi.org/10.1088/0004-6256/139/2/447},
   DOI={10.1088/0004-6256/139/2/447},
   number={2},
   journal={\aj},
   publisher={American Astronomical Society},
   author={Hunter, Deidre A. and Elmegreen, Bruce G. and Ludka, Bonnie C.},
   year={2010},
   month=jan, pages={447} }

@ARTICLE{Iorio_2017_data,
       author = {{Iorio}, G. and {Fraternali}, F. and {Nipoti}, C. and {Di Teodoro}, E. and {Read}, J.~I. and {Battaglia}, G.},
        title = "{LITTLE THINGS in 3D: robust determination of the circular velocity of dwarf irregular galaxies}",
      journal = {\mnras},
         year = {2017},
        month = {apr},
       volume = {466},
       number = {4},
        pages = {4159-4192},
          doi = {10.1093/mnras/stw3285},
archivePrefix = {arXiv},
       eprint = {1611.03865},
 primaryClass = {astro-ph.GA},
       adsurl = {https://ui.adsabs.harvard.edu/abs/2017MNRAS.466.4159I}
}

@ARTICLE{Leroy_2008,
       author = {{Leroy}, Adam K. and {Walter}, Fabian and {Brinks}, Elias and {Bigiel}, Frank and {de Blok}, W.~J.~G. and {Madore}, Barry and {Thornley}, M.~D.},
        title = "{The Star Formation Efficiency in Nearby Galaxies: Measuring Where Gas Forms Stars Effectively}",
      journal = {\aj},
         year = {2008},
        month = dec,
       volume = {136},
       number = {6},
        pages = {2782-2845},
          doi = {10.1088/0004-6256/136/6/2782},
archivePrefix = {arXiv},
       eprint = {0810.2556},
 primaryClass = {astro-ph},
       adsurl = {https://ui.adsabs.harvard.edu/abs/2008AJ....136.2782L}
}

@ARTICLE{Hao_2011,
       author = {{Hao}, Cai-Na and {Kennicutt}, Robert C. and {Johnson}, Benjamin D. and {Calzetti}, Daniela and {Dale}, Daniel A. and {Moustakas}, John},
        title = "{Dust-corrected Star Formation Rates of Galaxies. II. Combinations of Ultraviolet and Infrared Tracers}",
      journal = {\apj},
         year = {2011},
        month = nov,
       volume = {741},
       number = {2},
          eid = {124},
        pages = {124},
          doi = {10.1088/0004-637X/741/2/124},
archivePrefix = {arXiv},
       eprint = {1108.2837},
 primaryClass = {astro-ph.CO},
       adsurl = {https://ui.adsabs.harvard.edu/abs/2011ApJ...741..124H}
}

@ARTICLE{THINGS_2008,
       author = {{Walter}, Fabian and {Brinks}, Elias and {de Blok}, W.~J.~G. and {Bigiel}, Frank and {Kennicutt}, Jr., Robert C. and {Thornley}, Michele D. and {Leroy}, Adam},
        title = "{THINGS: The H I Nearby Galaxy Survey}",
      journal = {\aj},
         year = 2008,
        month = dec,
       volume = {136},
       number = {6},
        pages = {2563-2647},
          doi = {10.1088/0004-6256/136/6/2563},
archivePrefix = {arXiv},
       eprint = {0810.2125},
 primaryClass = {astro-ph},
       adsurl = {https://ui.adsabs.harvard.edu/abs/2008AJ....136.2563W}
}

@ARTICLE{2012_LITTLETHINGS,
       author = {{Hunter}, Deidre A. and {Ficut-Vicas}, Dana and {Ashley}, Trisha and {Brinks}, Elias and {Cigan}, Phil and {Elmegreen}, Bruce G. and {Heesen}, Volker and {Herrmann}, Kimberly A. and {Johnson}, Megan and {Oh}, Se-Heon and {Rupen}, Michael P. and {Schruba}, Andreas and {Simpson}, Caroline E. and {Walter}, Fabian and {Westpfahl}, David J. and {Young}, Lisa M. and {Zhang}, Hong-Xin},
        title = "{Little Things}",
      journal = {\aj},
         year = {2012},
        month = nov,
       volume = {144},
       number = {5},
          eid = {134},
        pages = {134},
          doi = {10.1088/0004-6256/144/5/134},
archivePrefix = {arXiv},
       eprint = {1208.5834},
 primaryClass = {astro-ph.GA},
       adsurl = {https://ui.adsabs.harvard.edu/abs/2012AJ....144..134H}
}

@ARTICLE{2011_Heald_HALOGAS,
       author = {{Heald}, G. and {J{\'o}zsa}, G. and {Serra}, P. and {Zschaechner}, L. and {Rand}, R. and {Fraternali}, F. and {Oosterloo}, T. and {Walterbos}, R. and {J{\"u}tte}, E. and {Gentile}, G.},
        title = "{The Westerbork Hydrogen Accretion in LOcal GAlaxieS (HALOGAS) survey. I. Survey description and pilot observations}",
      journal = {\aap},
         year = 2011,
        month = feb,
       volume = {526},
          eid = {A118},
        pages = {A118},
          doi = {10.1051/0004-6361/201015938},
archivePrefix = {arXiv},
       eprint = {1012.0816},
 primaryClass = {astro-ph.CO},
       adsurl = {https://ui.adsabs.harvard.edu/abs/2011A&A...526A.118H}
}

@ARTICLE{1984_u_corr,
       author = {{Burstein}, D. and {Heiles}, C.},
        title = "{Reddening estimates for galaxies in the second reference catalog and the Uppsala General Catalog.}",
      journal = {\apjs},
         year = {1984},
        month = jan,
       volume = {54},
        pages = {33-79},
          doi = {10.1086/190918},
       adsurl = {https://ui.adsabs.harvard.edu/abs/1984ApJS...54...33B}
}

@ARTICLE{2007_Wyder,
       author = {{Wyder}, Ted K. and {Martin}, D. Christopher and {Schiminovich}, David and {Seibert}, Mark and {Budav{\'a}ri}, Tam{\'a}s and {Treyer}, Marie A. and {Barlow}, Tom A. and {Forster}, Karl and {Friedman}, Peter G. and {Morrissey}, Patrick and {Neff}, Susan G. and {Small}, Todd and {Bianchi}, Luciana and {Donas}, Jos{\'e} and {Heckman}, Timothy M. and {Lee}, Young-Wook and {Madore}, Barry F. and {Milliard}, Bruno and {Rich}, R. Michael and {Szalay}, Alex S. and {Welsh}, Barry Y. and {Yi}, Sukyoung K.},
        title = "{The UV-Optical Galaxy Color-Magnitude Diagram. I. Basic Properties}",
      journal = {\apjs},
         year = {2007},
        month = dec,
       volume = {173},
       number = {2},
        pages = {293-314},
          doi = {10.1086/521402},
archivePrefix = {arXiv},
       eprint = {0706.3938},
 primaryClass = {astro-ph},
       adsurl = {https://ui.adsabs.harvard.edu/abs/2007ApJS..173..293W}
}

@ARTICLE{2015_barolo,
       author = {{Di Teodoro}, E.~M. and {Fraternali}, F.},
        title = "{$^{3D}$ BAROLO: a new 3D algorithm to derive rotation curves of galaxies}",
      journal = {\mnras},
         year = {2015},
        month = aug,
       volume = {451},
       number = {3},
        pages = {3021-3033},
          doi = {10.1093/mnras/stv1213},
archivePrefix = {arXiv},
       eprint = {1505.07834},
 primaryClass = {astro-ph.GA},
       adsurl = {https://ui.adsabs.harvard.edu/abs/2015MNRAS.451.3021D}
}

@article{2025_hunter_timescales,
doi = {10.3847/1538-3881/ae623f},
url = {https://doi.org/10.3847/1538-3881/ae623f},
year = {2026},
month = {may},
publisher = {The American Astronomical Society},
volume = {171},
number = {6},
pages = {358},
author = {Hunter, Laura Congreve and van Zee, Liese and Cohen, Roger E. and McQuinn, Kristen B. W. and Markham, Madison and Kader, Justin A. and Gault, Lexi and Dolphin, Andrew E.},
title = {Observed Timescales of Stellar Feedback in Star-forming, Low-mass Galaxies},
journal = {\aj}
}

@ARTICLE{2000_turbulence_crossingtime_elmegreen,
       author = {{Elmegreen}, Bruce G.},
        title = "{Star Formation in a Crossing Time}",
      journal = {\apj},
         year = {2000},
        month = feb,
       volume = {530},
       number = {1},
        pages = {277-281},
          doi = {10.1086/308361},
archivePrefix = {arXiv},
       eprint = {astro-ph/9911172},
 primaryClass = {astro-ph},
       adsurl = {https://ui.adsabs.harvard.edu/abs/2000ApJ...530..277E}
}

@article{Stilp_2013_drivers_of_turbulence,
        title={DRIVERS OF H I TURBULENCE IN DWARF GALAXIES},
        volume={773},
        ISSN={1538-4357},
        url={http://dx.doi.org/10.1088/0004-637X/773/2/88},
        DOI={10.1088/0004-637x/773/2/88},
        number={2},
        journal={\apj},
        publisher={American Astronomical Society},
        author={Stilp, Adrienne M. and Dalcanton, Julianne J. and Skillman, Evan and Warren, Steven R. and Ott, Jürgen and Koribalski, Bärbel},
        year={2013},
        month={jul}, 
        pages={88} }

@article{Mac_Low_2004,
        title={Control of star formation by supersonic turbulence},
        volume={76},
        ISSN={1539-0756},
        url={http://dx.doi.org/10.1103/RevModPhys.76.125},
        DOI={10.1103/revmodphys.76.125},
        number={1},
        journal={Rev. Mod. Phys.},
        publisher={American Physical Society (APS)},
        author={Mac Low, Mordecai-Mark and Klessen, Ralf S.},
        year={2004},
        month=jan, pages={125} }

@ARTICLE{1941Kolmogorov,
        author = {{Kolmogorov}, A.},
        title = "{The Local Structure of Turbulence in Incompressible Viscous Fluid for Very Large Reynolds' Numbers}",
        journal = {Akademiia Nauk SSSR Doklady},
        year = 1941,
        month = jan,
        volume = {30},
        pages = {301-305},
        adsurl = {https://ui.adsabs.harvard.edu/abs/1941DoSSR..30..301K}
        }

@article{Hunter_2021_correlation,
        doi = {10.3847/1538-3881/abe1c0},
        url = {https://dx.doi.org/10.3847/1538-3881/abe1c0},
        year = {2021},
        month = {mar},
        publisher = {The American Astronomical Society},
        volume = {161},
        number = {4},
        pages = {175},
        author = {Hunter, Deidre A. and Elmegreen, Bruce G. and Archer, Haylee and Simpson, Caroline E. and Cigan, Phil},
        title = {A Search for Correlations between Turbulence and Star Formation in LITTLE THINGS Dwarf Irregular Galaxies},
        journal = {\aj}
        }

@article{Stone_1998,
        doi = {10.1086/311718},
        url = {https://doi.org/10.1086/311718},
        year = {1998},
        month = {oct},
        publisher = {},
        volume = {508},
        number = {1},
        pages = {L99},
        author = {Stone, James M. and Ostriker, Eve C. and Gammie, Charles F.},
        title = {Dissipation in Compressible Magnetohydrodynamic Turbulence},
        journal = {\apj}
        }

@article{Padoan_1999,
        doi = {10.1086/307956},
        url = {https://doi.org/10.1086/307956},
        year = {1999},
        month = {nov},
        volume = {526},
        number = {1},
        pages = {279},
        author = {Padoan, Paolo and Nordlund, {\aa}ke},
        title = {A Super-Alfvénic Model of Dark Clouds},
        journal = {\apj}
        }

@article{MacLow_1999,
        doi = {10.1086/307784},
        url = {https://doi.org/10.1086/307784},
        year = {1999},
        month = {oct},
        publisher = {},
        volume = {524},
        number = {1},
        pages = {169},
        author = {Mac Low, Mordecai-Mark},
        title = {The Energy Dissipation Rate of Supersonic, Magnetohydrodynamic Turbulence in Molecular Clouds},
        journal = {\apj}
        }

@ARTICLE{Dutta_2009,
        author = {{Dutta}, Prasun and {Begum}, Ayesha and {Bharadwaj}, Somnath and {Chengalur}, Jayaram N.},
        title = "{A study of interstellar Fmedium of dwarf galaxies using HI power spectrum analysis}",
        journal = {\mnras},
        year = 2009,
        month = sep,
        volume = {398},
        number = {2},
        pages = {887-897},
        doi = {10.1111/j.1365-2966.2009.15105.x},
        archivePrefix = {arXiv},
        eprint = {0905.1756},
        primaryClass = {astro-ph.GA},
        adsurl = {https://ui.adsabs.harvard.edu/abs/2009MNRAS.398..887D}
        }

@ARTICLE{2005_willet_dwarf_vs_kolmo,
                author = {{Willett}, Kyle W. and {Elmegreen}, Bruce G. and {Hunter}, Deidre A.},
                title = "{Power Spectra in V band and H{\ensuremath{\alpha}} of Nine Irregular Galaxies}",
                journal = {\aj},
                year = {2005},
                month = may,
                volume = {129},
                number = {5},
                pages = {2186-2196},
                doi = {10.1086/429678},
                archivePrefix = {arXiv},
                eprint = {astro-ph/0503295},
                primaryClass = {astro-ph},
                adsurl = {https://ui.adsabs.harvard.edu/abs/2005AJ....129.2186W}
                }

@article{Tamburro_2009,
            doi = {10.1088/0004-6256/137/5/4424},
            url = {https://dx.doi.org/10.1088/0004-6256/137/5/4424},
            year = {2009},
            month = {apr},
            publisher = {The American Astronomical Society},
            volume = {137},
            number = {5},
            pages = {4424},
            author = {Tamburro, D. and Rix, H.-W. and Leroy, A. K. and Low, M.-M. Mac and Walter, F. and Kennicutt, R. C. and Brinks, E. and de Blok, W. J. G.},
            title = {WHAT IS DRIVING THE H i VELOCITY DISPERSION?},
            journal = {\aj}
            }

@ARTICLE{2015_SN_energy_kim_ostriker,
            author = {{Kim}, Chang-Goo and {Ostriker}, Eve C.},
            title = "{Momentum Injection by Supernovae in the Interstellar Medium}",
            journal = {\apj},
            year = {2015},
            month = {apr},
            volume = {802},
            number = {2},
            eid = {99},
            pages = {99},
            doi = {10.1088/0004-637X/802/2/99},
            archivePrefix = {arXiv},
            eprint = {1410.1537},
            primaryClass = {astro-ph.GA},
            adsurl = {https://ui.adsabs.harvard.edu/abs/2015ApJ...802...99K}
            }

@ARTICLE{2014_Saury,
            author = {{Saury}, E. and {Miville-Desch{\^e}nes}, M.-A. and {Hennebelle}, P. and {Audit}, E. and {Schmidt}, W.},
            title = "{The structure of the thermally bistable and turbulent atomic gas in the local interstellar medium}",
            journal = {\aap},
            year = 2014,
            month = jul,
            volume = {567},
            eid = {A16},
            pages = {A16},
            doi = {10.1051/0004-6361/201321113},
            archivePrefix = {arXiv},
            eprint = {1301.3446},
            primaryClass = {astro-ph.GA},
            adsurl = {https://ui.adsabs.harvard.edu/abs/2014A&A...567A..16S}
            }

@ARTICLE{1977_McKee_Ostriker,
            author = {{McKee}, C.~F. and {Ostriker}, J.~P.},
            title = "{A theory of the interstellar medium: three components regulated by supernova explosions in an inhomogeneous substrate.}",
            journal = {\apj},
            year = 1977,
            month = nov,
            volume = {218},
            pages = {148-169},
            doi = {10.1086/155667},
            adsurl = {https://ui.adsabs.harvard.edu/abs/1977ApJ...218..148M}
            }

@INPROCEEDINGS{1987_kulkarni_heiles,
            author = {{Kulkarni}, Shrinivas R. and {Heiles}, Carl},
            title = "{The Atomic Component}",
            booktitle = {Interstellar Processes},
            year = 1987,
            editor = {{Hollenbach}, David J. and {Thronson}, Jr., Harley A.},
            volume = {134},
            month = jan,
            pages = {87},
            doi = {10.1007/978-94-009-3861-8_5},
            adsurl = {https://ui.adsabs.harvard.edu/abs/1987ASSL..134...87K}
            }

@article{Roy_2013,
            author = {Roy, Nirupam and Kanekar, Nissim and Chengalur, Jayaram N.},
            title = {The temperature of the diffuse H i in the Milky Way - II. Gaussian decomposition of the H i-21 cm absorption spectra},
            journal = {\mnras},
            volume = {436},
            number = {3},
            pages = {2366-2385},
            year = {2013},
            month = {10},
            issn = {0035-8711},
            doi = {10.1093/mnras/stt1746},
            url = {https://doi.org/10.1093/mnras/stt1746},
            eprint = {https://academic.oup.com/mnras/article-pdf/436/3/2366/4084982/stt1746.pdf},
            }

@article{Kalberla_2018,
            title={Properties of cold and warm H I gas phases derived from a Gaussian decomposition of HI4PI data},
            volume={619},
            ISSN={1432-0746},
            url={http://dx.doi.org/10.1051/0004-6361/201833146},
            DOI={10.1051/0004-6361/201833146},
            journal={\aap},
            publisher={EDP Sciences},
            author={Kalberla, P. M. W. and Haud, U.},
            year={2018},
            month=nov, pages={A58} }

@ARTICLE{2003_Heiles_Troland_WNM,
            author = {{Heiles}, Carl and {Troland}, T.~H.},
            title = "{The Millennium Arecibo 21 Centimeter Absorption-Line Survey. II. Properties of the Warm and Cold Neutral Media}",
            journal = {\apj},
            year = {2003},
            month = {apr},
            volume = {586},
            number = {2},
            pages = {1067-1093},
            doi = {10.1086/367828},
            archivePrefix = {arXiv},
            eprint = {astro-ph/0207105},
            primaryClass = {astro-ph},
            adsurl = {https://ui.adsabs.harvard.edu/abs/2003ApJ...586.1067H}
            }

@ARTICLE{2013_pineda,
            author = {{Pineda}, J.~L. and {Langer}, W.~D. and {Velusamy}, T. and {Goldsmith}, P.~F.},
            title = "{A Herschel [C ii] Galactic plane survey. I. The global distribution of ISM gas components}",
            journal = {\aap},
            year = {2013},
            month = {jun},
            volume = {554},
            eid = {A103},
            pages = {A103},
            doi = {10.1051/0004-6361/201321188},
            archivePrefix = {arXiv},
            eprint = {1304.7770},
            primaryClass = {astro-ph.GA},
            adsurl = {https://ui.adsabs.harvard.edu/abs/2013A&A...554A.103P}
            }

@article{Dickey_2000,
            doi = {10.1086/308953},
            url = {https://doi.org/10.1086/308953},
            year = {2000},
            month = {jun},
            publisher = {},
            volume = {536},
            number = {2},
            pages = {756},
            author = {Dickey, John M. and Mebold, Ulrich and Stanimirovic, Snezana and Staveley-Smith, Lister},
            title = {Cold Atomic Gas in the Small Magellanic Cloud},
            journal = {\apj}
            }

@ARTICLE{2000-zimmer,
            author = {{Marx-Zimmer}, M. and {Herbstmeier}, U. and {Dickey}, J.~M. and {Zimmer}, F. and {Staveley-Smith}, L. and {Mebold}, U.},
            title = "{A study of the cool gas in the Large Magellanic Cloud. I. Properties of the cool atomic phase - a third H i absorption survey}",
            journal = {\aap},
            year = 2000,
            month = feb,
            volume = {354},
            pages = {787-801},
            adsurl = {https://ui.adsabs.harvard.edu/abs/2000A&A...354..787M}
            }

@ARTICLE{wolfire_2003_temperatures,
            author = {{Wolfire}, Mark G. and {McKee}, Christopher F. and {Hollenbach}, David and {Tielens}, A.~G.~G.~M.},
            title = "{Neutral Atomic Phases of the Interstellar Medium in the Galaxy}",
            journal = {\apj},
            year = 2003,
            month = apr,
            volume = {587},
            number = {1},
            pages = {278-311},
            doi = {10.1086/368016},
            archivePrefix = {arXiv},
            eprint = {astro-ph/0207098},
            primaryClass = {astro-ph},
            adsurl = {https://ui.adsabs.harvard.edu/abs/2003ApJ...587..278W}
            }

@article{de_Blok_2008,
            doi = {10.1088/0004-6256/136/6/2648},
            url = {https://doi.org/10.1088/0004-6256/136/6/2648},
            year = {2008},
            month = {nov},
            publisher = {The American Astronomical Society},
            volume = {136},
            number = {6},
            pages = {2648},
            author = {de Blok, W. J. G. and Walter, F. and Brinks, E. and Trachternach, C. and Oh, S-H. and Kennicutt, R. C.},
            title = {HIGH-RESOLUTION ROTATION CURVES AND GALAXY MASS MODELS FROM THINGS},
            journal = {\aj}
            }

@ARTICLE{2016_frank,
            author = {{Frank}, B.~S. and {de Blok}, W.~J.~G. and {Walter}, F. and {Leroy}, A. and {Carignan}, C.},
            title = "{The Impact of Molecular Gas on Mass Models of Nearby Galaxies}",
            journal = {\aj},
            year = 2016,
            month = apr,
            volume = {151},
            number = {4},
            eid = {94},
            pages = {94},
            doi = {10.3847/0004-6256/151/4/94},
            archivePrefix = {arXiv},
            eprint = {1512.01367},
            primaryClass = {astro-ph.GA},
            adsurl = {https://ui.adsabs.harvard.edu/abs/2016AJ....151...94F}
            }

@ARTICLE{2005_read_gilmore,
            author = {{Read}, J.~I. and {Gilmore}, G.},
            title = "{Mass loss from dwarf spheroidal galaxies: the origins of shallow dark matter cores and exponential surface brightness profiles}",
            journal = {\mnras},
            year = 2005,
            month = jan,
            volume = {356},
            number = {1},
            pages = {107-124},
            doi = {10.1111/j.1365-2966.2004.08424.x},
            archivePrefix = {arXiv},
            eprint = {astro-ph/0409565},
            primaryClass = {astro-ph},
            adsurl = {https://ui.adsabs.harvard.edu/abs/2005MNRAS.356..107R}
            }

@ARTICLE{2014_Pontzen_Governato,
            author = {{Pontzen}, Andrew and {Governato}, Fabio},
            title = "{Cold dark matter heats up}",
            journal = {\nat},
            year = 2014,
            month = feb,
            volume = {506},
            number = {7487},
            pages = {171-178},
            doi = {10.1038/nature12953},
            archivePrefix = {arXiv},
            eprint = {1402.1764},
            primaryClass = {astro-ph.CO},
            adsurl = {https://ui.adsabs.harvard.edu/abs/2014Natur.506..171P}
            }

@article{Jameson_2019,
            title={An ATCA Survey of H i Absorption in the Magellanic Clouds. I. H i Gas Temperature Measurements in the Small Magellanic Cloud},
            volume={244},
            ISSN={1538-4365},
            url={http://dx.doi.org/10.3847/1538-4365/ab3576},
            DOI={10.3847/1538-4365/ab3576},
            number={1},
            journal={\apjs},
            publisher={American Astronomical Society},
            author={Jameson, Katherine E. and McClure-Griffiths, N. M. and Liu, Boyang and             Dickey, John M. and Staveley-Smith, Lister and Stanimirović, Snežana and Dempsey,          James and Dawson, J. R. and Dénes, Helga and Bolatto, Alberto D. and Li, Di and            Wong, Tony},
            year={2019},
            month=sep, 
            pages={7}}

@article{Chen_2025,
            title={A Neutral Hydrogen Absorption Study of Cold Gas in the Outskirts of the          Magellanic Clouds Using the GASKAP-H <scp>i</scp> Survey},
            volume={169},
            ISSN={1538-3881},
            url={http://dx.doi.org/10.3847/1538-3881/adc67c},
            DOI={10.3847/1538-3881/adc67c},
            number={5},
            journal={\aj},
            publisher={American Astronomical Society},
            author={Chen, Hongxing and Stanimirović, Snežana and Pingel, Nickolas M. and            Dempsey, James and Buckland-Willis, Frances and Clark, Susan E. and Dénes, Helga        and Dickey, John M. and Gibson, Steven and Jameson, Katherine and Kemp, Ian and          Leahy, Denis and Lee, Min-Young and Lynn, Callum and Ma, Yik Ki and McClure-      Griffiths, N. M. and Murray, Claire E. and Nguyen, Hiep and Uscanga, Lucero and       van Loon, Jacco Th. and Vázquez-Semadeni, Enrique},
            year={2025},
            month=apr, 
            pages={284}}

@article{Young_2003_WNM_dIrrs,
        title={Star Formation and the Interstellar Medium in Four Dwarf Irregular Galaxies},
        volume={592},
        ISSN={1538-4357},
        url={http://dx.doi.org/10.1086/375581},
        DOI={10.1086/375581},
        number={1},
        journal={\apj},
        publisher={American Astronomical Society},
        author={Young, L. M. and van Zee, L. and Lo, K. Y. and Dohm‐Palmer, R. C. and Beierle, Michelle E.},
        year={2003},
        month={jul}, 
        pages={111}
        }

@article{Hunter_2021,
        doi = {10.3847/1538-3881/abd089},
        url = {https://dx.doi.org/10.3847/1538-3881/abd089},
        year = {2021},
        month = {jan},
        publisher = {The American Astronomical Society},
        volume = {161},
        number = {2},
        pages = {71},
        author = {Hunter, Deidre A. and Elmegreen, Bruce G. and Goldberger, Esther and Taylor, Hannah and Ermakov, Anton I. and Herrmann, Kimberly A. and Oh, Se-Heon and Malko, Bradley and Barandi, Brian and Jundt, Ryan},
        title = {Relationships between the Stellar, Gaseous, and Star Formation Disks in LITTLE THINGS Dwarf Irregular Galaxies: Indirect Evidence for Substantial Fractions of Dark Molecular Gas},
        journal = {\aj}
        }

@ARTICLE{hunter2024interstellarmediumdwarfirregular,
       author = {{Hunter}, Deidre A. and {Elmegreen}, Bruce G. and {Madden}, Suzanne C.},
        title = "{The Interstellar Medium in Dwarf Irregular Galaxies}",
      journal = {\araa},
         year = 2024,
        month = sep,
       volume = {62},
       number = {1},
        pages = {113-155},
          doi = {10.1146/annurev-astro-052722-104109},
archivePrefix = {arXiv},
       eprint = {2402.17004},
 primaryClass = {astro-ph.GA},
       adsurl = {https://ui.adsabs.harvard.edu/abs/2024ARA&A..62..113H}
}

@article{Pokhrel_2020,
            doi = {10.3847/1538-3881/ab9bfa},
            url = {https://dx.doi.org/10.3847/1538-3881/ab9bfa},
            year = {2020},
            month = {jul},
            publisher = {The American Astronomical Society},
            volume = {160},
            number = {2},
            pages = {66},
            author = {Pokhrel, Nau Raj and Simpson, Caroline E. and Bagetakos, Ioannis},
            title = {A Catalog of Holes and Shells in the Interstellar Medium of the LITTLE THINGS Dwarf Galaxies},
            journal = {\aj}
            }

@article{Bacchini_2019a,
            title={Volumetric star formation laws of disc galaxies},
            volume={622},
            ISSN={1432-0746},
            url={http://dx.doi.org/10.1051/0004-6361/201834382},
            DOI={10.1051/0004-6361/201834382},
            journal={\aap},
            publisher={EDP Sciences},
            author={Bacchini, Cecilia and Fraternali, Filippo and Iorio, Giuliano and Pezzulli, Gabriele},
            year={2019},
            month=jan, pages={A64} }

@ARTICLE{2016_read_DM_core,
            author = {{Read}, J.~I. and {Agertz}, O. and {Collins}, M.~L.~M.},
            title = "{Dark matter cores all the way down}",
            journal = {\mnras},
            year = {2016},
            month = {jul},
            volume = {459},
            number = {3},
            pages = {2573-2590},
            doi = {10.1093/mnras/stw713},
            archivePrefix = {arXiv},
            eprint = {1508.04143},
            primaryClass = {astro-ph.GA},
            adsurl = {https://ui.adsabs.harvard.edu/abs/2016MNRAS.459.2573R}
            }

@ARTICLE{2008_boomsma,
            author = {{Boomsma}, R. and {Oosterloo}, T.~A. and {Fraternali}, F. and {van der Hulst}, J.~M. and {Sancisi}, R.},
            title = "{HI holes and high-velocity clouds in the spiral galaxy NGC 6946}",
            journal = {\aap},
            year = 2008,
            month = nov,
            volume = {490},
            number = {2},
            pages = {555-570},
            doi = {10.1051/0004-6361:200810120},
            archivePrefix = {arXiv},
            eprint = {0807.3339},
            primaryClass = {astro-ph},
            adsurl = {https://ui.adsabs.harvard.edu/abs/2008A&A...490..555B}
            }

@ARTICLE{1989_SNR_blowout_maclow,
            author = {{Mac Low}, Mordecai-Mark and {McCray}, Richard and {Norman}, Michael L.},
            title = "{Superbubble Blowout Dynamics}",
            journal = {\apj},
            year = 1989,
            month = feb,
            volume = {337},
            pages = {141},
            doi = {10.1086/167094},
            adsurl = {https://ui.adsabs.harvard.edu/abs/1989ApJ...337..141M}
            }

@article{Fierlinger_SNR,
            author = {Fierlinger, Katharina M. and Burkert, Andreas and Ntormousi, Evangelia and Fierlinger, Peter and Schartmann, Marc and Ballone, Alessandro and Krause, Martin G. H. and Diehl, Roland},
            title = {Stellar feedback efficiencies: supernovae versus stellar winds},
            journal = {\mnras},
            volume = {456},
            number = {1},
            pages = {710-730},
            year = {2015},
            month = {12},
            issn = {0035-8711},
            doi = {10.1093/mnras/stv2699},
            url = {https://doi.org/10.1093/mnras/stv2699},
            eprint = {https://academic.oup.com/mnras/article-pdf/456/1/710/3707016/stv2699.pdf},
            }

@article{LauBonnell2025,
            author = {Lau, Cheryl S C and Bonnell, Ian A},
            title = {Semi-confined supernova feedback in H ii region bubbles},
            journal = {\mnras},
            volume = {540},
            number = {1},
            pages = {1124-1143},
            year = {2025},
            month = {06},
            issn = {0035-8711},
            doi = {10.1093/mnras/staf756},
            url = {https://doi.org/10.1093/mnras/staf756},
            eprint = {https://academic.oup.com/mnras/article-pdf/540/1/1124/63129601/staf756.pdf},
            }

@ARTICLE{2025_maltsev,
            author = {{Maltsev}, K. and {Schneider}, F.~R.~N. and {Mandel}, I. and {M{\"u}ller}, B. and {Heger}, A. and {R{\"o}pke}, F.~K. and {Laplace}, E.},
            title = "{Explodability criteria for the neutrino-driven supernova mechanism}",
            journal = {\aap},
            year = 2025,
            month = aug,
            volume = {700},
            eid = {A20},
            pages = {A20},
            doi = {10.1051/0004-6361/202554931},
            archivePrefix = {arXiv},
            eprint = {2503.23856},
            primaryClass = {astro-ph.SR},
            adsurl = {https://ui.adsabs.harvard.edu/abs/2025A&A...700A..20M}
            }

@article{McQuinn_2015,
            doi = {10.1088/0004-637X/808/2/109},
            url = {https://dx.doi.org/10.1088/0004-637X/808/2/109},
            year = {2015},
            month = {jul},
            publisher = {The American Astronomical Society},
            volume = {808},
            number = {2},
            pages = {109},
            author = {McQuinn, Kristen B. W. and Skillman, Evan D. and Dolphin, Andrew E. and Mitchell, Noah P.},
            title = {CALIBRATING UV STAR FORMATION RATES FOR DWARF GALAXIES FROM STARBIRDS*},
            journal = {\apj}
            }

@ARTICLE{2023_Popesso,
       author = {{Popesso}, P. and {Concas}, A. and {Cresci}, G. and {Belli}, S. and {Rodighiero}, G. and {Inami}, H. and {Dickinson}, M. and {Ilbert}, O. and {Pannella}, M. and {Elbaz}, D.},
        title = "{The main sequence of star-forming galaxies across cosmic times}",
      journal = {\mnras},
         year = 2023,
        month = feb,
       volume = {519},
       number = {1},
        pages = {1526-1544},
          doi = {10.1093/mnras/stac3214},
archivePrefix = {arXiv},
       eprint = {2203.10487},
 primaryClass = {astro-ph.GA},
       adsurl = {https://ui.adsabs.harvard.edu/abs/2023MNRAS.519.1526P}
}

@ARTICLE{Berg2022,
       author = {{Berg}, Danielle A. and {James}, Bethan L. and {King}, Teagan and {McDonald}, Meaghan and {Chen}, Zuyi and {Chisholm}, John and {Heckman}, Timothy and {Martin}, Crystal L. and {Stark}, Dan P. and {Aloisi}, Alessandra and {Amor{\'\i}n}, Ricardo O. and {Arellano-C{\'o}rdova}, Karla Z. and {Bayliss}, Matthew and {Bordoloi}, Rongmon and {Brinchmann}, Jarle and {Charlot}, St{\'e}phane and {Chevallard}, Jacopo and {Clark}, Ilyse and {Erb}, Dawn K. and {Feltre}, Anna and {Gronke}, Max and {Hayes}, Matthew and {Henry}, Alaina and {Hernandez}, Svea and {Jaskot}, Anne and {Jones}, Tucker and {Kewley}, Lisa J. and {Kumari}, Nimisha and {Leitherer}, Claus and {Llerena}, Mario and {Maseda}, Michael and {Mingozzi}, Matilde and {Nanayakkara}, Themiya and {Ouchi}, Masami and {Plat}, Adele and {Pogge}, Richard W. and {Ravindranath}, Swara and {Rigby}, Jane R. and {Sanders}, Ryan and {Scarlata}, Claudia and {Senchyna}, Peter and {Skillman}, Evan D. and {Steidel}, Charles C. and {Strom}, Allison L. and {Sugahara}, Yuma and {Wilkins}, Stephen M. and {Wofford}, Aida and {Xu}, Xinfeng and {Classy Team}},
        title = "{The COS Legacy Archive Spectroscopy Survey (CLASSY) Treasury Atlas}",
      journal = {\apjs},
         year = 2022,
        month = aug,
       volume = {261},
       number = {2},
          eid = {31},
        pages = {31},
          doi = {10.3847/1538-4365/ac6c03},
archivePrefix = {arXiv},
       eprint = {2203.07357},
 primaryClass = {astro-ph.GA},
       adsurl = {https://ui.adsabs.harvard.edu/abs/2022ApJS..261...31B}
}

@ARTICLE{2016_padoan,
       author = {{Padoan}, Paolo and {Pan}, Liubin and {Haugb{\o}lle}, Troels and {Nordlund}, {\r{A}}ke},
        title = "{Supernova Driving. I. The Origin of Molecular Cloud Turbulence}",
      journal = {\apj},
         year = 2016,
        month = may,
       volume = {822},
       number = {1},
          eid = {11},
        pages = {11},
          doi = {10.3847/0004-637X/822/1/11},
archivePrefix = {arXiv},
       eprint = {1509.04663},
 primaryClass = {astro-ph.GA},
       adsurl = {https://ui.adsabs.harvard.edu/abs/2016ApJ...822...11P}
}

@ARTICLE{2016_Pan,
       author = {{Pan}, Liubin and {Padoan}, Paolo and {Haugb{\o}lle}, Troels and {Nordlund}, {\r{A}}ke},
        title = "{Supernova Driving. II. Compressive Ratio in Molecular-cloud Turbulence}",
      journal = {\apj},
         year = 2016,
        month = jul,
       volume = {825},
       number = {1},
          eid = {30},
        pages = {30},
          doi = {10.3847/0004-637X/825/1/30},
archivePrefix = {arXiv},
       eprint = {1510.04742},
 primaryClass = {astro-ph.GA},
       adsurl = {https://ui.adsabs.harvard.edu/abs/2016ApJ...825...30P}
}

@ARTICLE{2017_Orkisz,
       author = {{Orkisz}, Jan H. and {Pety}, J{\'e}r{\^o}me and {Gerin}, Maryvonne and {Bron}, Emeric and {Guzm{\'a}n}, Viviana V. and {Bardeau}, S{\'e}bastien and {Goicoechea}, Javier R. and {Gratier}, Pierre and {Le Petit}, Franck and {Levrier}, Fran{\c{c}}ois and {Liszt}, Harvey and {{\"O}berg}, Karin and {Peretto}, Nicolas and {Roueff}, Evelyne and {Sievers}, Albrecht and {Tremblin}, Pascal},
        title = "{Turbulence and star formation efficiency in molecular clouds: solenoidal versus compressive motions in Orion B}",
      journal = {\aap},
         year = 2017,
        month = mar,
       volume = {599},
          eid = {A99},
        pages = {A99},
          doi = {10.1051/0004-6361/201629220},
archivePrefix = {arXiv},
       eprint = {1701.00962},
 primaryClass = {astro-ph.GA},
       adsurl = {https://ui.adsabs.harvard.edu/abs/2017A&A...599A..99O}
}

@ARTICLE{betancourt2013hamiltonianmontecarlohierarchical,
       author = {{Betancourt}, M.~J. and {Girolami}, Mark},
        title = "{Hamiltonian Monte Carlo for Hierarchical Models}",
      journal = {arXiv e-prints},
         year = 2013,
        month = dec,
          eid = {arXiv:1312.0906},
        pages = {arXiv:1312.0906},
          doi = {10.48550/arXiv.1312.0906},
archivePrefix = {arXiv},
       eprint = {1312.0906},
 primaryClass = {stat.ME},
       adsurl = {https://ui.adsabs.harvard.edu/abs/2013arXiv1312.0906B}
}

@article{papaspiliopoulos,
author = {Papaspiliopoulos, Omiros and Roberts, Gareth},
year = {2003},
month = {01},
pages = {307-326},
title = {Non-Centered Parameterisations for Hierarchical Models and Data Augmentation},
volume = {7},
journal = {Bayesian Statistics}
}

@ARTICLE{2019_Marasco,
       author = {{Marasco}, A. and {Fraternali}, F. and {Heald}, G. and {de Blok}, W.~J.~G. and {Oosterloo}, T. and {Kamphuis}, P. and {J{\'o}zsa}, G.~I.~G. and {Vargas}, C.~J. and {Winkel}, B. and {Walterbos}, R.~A.~M. and {Dettmar}, R.~J. and {Juẗte}, E.},
        title = "{HALOGAS: the properties of extraplanar HI in disc galaxies}",
      journal = {\aap},
         year = 2019,
        month = nov,
       volume = {631},
          eid = {A50},
        pages = {A50},
          doi = {10.1051/0004-6361/201936338},
archivePrefix = {arXiv},
       eprint = {1909.04048},
 primaryClass = {astro-ph.GA},
       adsurl = {https://ui.adsabs.harvard.edu/abs/2019A&A...631A..50M}
}

@article{Ostriker_2011,
doi = {10.1088/0004-637X/731/1/41},
url = {https://doi.org/10.1088/0004-637X/731/1/41},
year = {2011},
month = {mar},
publisher = {The American Astronomical Society},
volume = {731},
number = {1},
pages = {41},
author = {Ostriker, Eve C. and Shetty, Rahul},
title = {MAXIMALLY STAR-FORMING GALACTIC DISKS. I. STARBURST REGULATION VIA FEEDBACK-DRIVEN TURBULENCE},
journal = {\apj}
}

@ARTICLE{2022_ostriker_kim,
       author = {{Ostriker}, Eve C. and {Kim}, Chang-Goo},
        title = "{Pressure-regulated, Feedback-modulated Star Formation in Disk Galaxies}",
      journal = {\apj},
         year = 2022,
        month = sep,
       volume = {936},
       number = {2},
          eid = {137},
        pages = {137},
          doi = {10.3847/1538-4357/ac7de2},
archivePrefix = {arXiv},
       eprint = {2206.00681},
 primaryClass = {astro-ph.GA},
       adsurl = {https://ui.adsabs.harvard.edu/abs/2022ApJ...936..137O}
}

@ARTICLE{2017MarascoFraternali,
       author = {{Marasco}, A. and {Fraternali}, F.},
        title = "{The Galactic fountain as an origin for the Smith Cloud}",
      journal = {\mnras},
         year = "2017",
        month = "Jan",
       volume = {464},
       number = {1},
        pages = {L100-L104},
          doi = {10.1093/mnrasl/slw195},
       adsurl = {https://ui.adsabs.harvard.edu/abs/2017MNRAS.464L.100M}
}

@ARTICLE{2023Marasco,
       author = {{Marasco}, A. and {Belfiore}, F. and {Cresci}, G. and {Lelli}, F. and {Venturi}, G. and {Hunt}, L.~K. and {Concas}, A. and {Marconi}, A. and {Mannucci}, F. and {Mingozzi}, M. and {McLeod}, A.~F. and {Kumari}, N. and {Carniani}, S. and {Vanzi}, L. and {Ginolfi}, M.},
        title = "{Shaken, but not expelled: Gentle baryonic feedback from nearby starburst dwarf galaxies}",
      journal = {\aap},
         year = 2023,
        month = feb,
       volume = {670},
          eid = {A92},
        pages = {A92},
          doi = {10.1051/0004-6361/202244895},
archivePrefix = {arXiv},
       eprint = {2209.02726},
 primaryClass = {astro-ph.GA},
       adsurl = {https://ui.adsabs.harvard.edu/abs/2023A&A...670A..92M}
}

@ARTICLE{2008Begum_a,
       author = {{Begum}, Ayesha and {Chengalur}, Jayaram N. and {Kennicutt}, Robert C. and {Karachentsev}, Igor D. and {Lee}, Janice C.},
        title = "{Life in the last lane: star formation and chemical evolution in an extremely gas rich dwarf}",
      journal = {\mnras},
         year = 2008,
        month = jan,
       volume = {383},
       number = {2},
        pages = {809-816},
          doi = {10.1111/j.1365-2966.2007.12592.x},
archivePrefix = {arXiv},
       eprint = {0711.1588},
 primaryClass = {astro-ph},
       adsurl = {https://ui.adsabs.harvard.edu/abs/2008MNRAS.383..809B}
}

@ARTICLE{2012Elson,
       author = {{Elson}, E.~C. and {de Blok}, W.~J.~G. and {Kraan-Korteweg}, R.~C.},
        title = "{Star Formation Models for the Dwarf Galaxies NGC 2915 and NGC 1705}",
      journal = {\aj},
         year = 2012,
        month = jan,
       volume = {143},
       number = {1},
          eid = {1},
        pages = {1},
          doi = {10.1088/0004-6256/143/1/1},
archivePrefix = {arXiv},
       eprint = {1110.1148},
 primaryClass = {astro-ph.CO},
       adsurl = {https://ui.adsabs.harvard.edu/abs/2012AJ....143....1E}
}

@ARTICLE{2017Namumba,
       author = {{Namumba}, B. and {Carignan}, C. and {Passmoor}, S. and {de Blok}, W.~J.~G.},
        title = "{H I kinematics, mass distribution and star formation threshold in NGC 6822, using the SKA pathfinder KAT-7}",
      journal = {\mnras},
         year = 2017,
        month = dec,
       volume = {472},
       number = {3},
        pages = {3761-3773},
          doi = {10.1093/mnras/stx2256},
archivePrefix = {arXiv},
       eprint = {1708.09447},
 primaryClass = {astro-ph.GA},
       adsurl = {https://ui.adsabs.harvard.edu/abs/2017MNRAS.472.3761N}
}

@ARTICLE{2018Watts,
       author = {{Watts}, Adam B. and {Meurer}, Gerhardt R. and {Lagos}, Claudia D.~P. and {Bruzzese}, Sarah M. and {Kroupa}, Pavel and {Jerabkova}, Tereza},
        title = "{Star formation in the outskirts of DDO 154: a top-light IMF in a nearly dormant disc}",
      journal = {\mnras},
         year = 2018,
        month = jul,
       volume = {477},
       number = {4},
        pages = {5554-5567},
          doi = {10.1093/mnras/sty1006},
archivePrefix = {arXiv},
       eprint = {1804.07072},
 primaryClass = {astro-ph.GA},
       adsurl = {https://ui.adsabs.harvard.edu/abs/2018MNRAS.477.5554W}
}

@ARTICLE{2018Murray,
       author = {{Murray}, Claire E. and {Stanimirovi{\'c}}, Sne{\v{z}}ana and
         {Goss}, W.~M. and {Heiles}, Carl and {Dickey}, John M. and
         {Babler}, Brian and {Kim}, Chang-Goo},
        title = "{The 21-SPONGE H I Absorption Line Survey. I. The Temperature of Galactic H I}",
      journal = {\apjs},
         year = 2018,
        month = oct,
       volume = {238},
       number = {2},
          eid = {14},
        pages = {14},
          doi = {10.3847/1538-4365/aad81a},
archivePrefix = {arXiv},
       eprint = {1806.06065},
 primaryClass = {astro-ph.GA},
       adsurl = {https://ui.adsabs.harvard.edu/abs/2018ApJS..238...14M}
}

@ARTICLE{1996Beck,
       author = {{Beck}, Rainer and {Brandenburg}, Axel and {Moss}, David and
         {Shukurov}, Anvar and {Sokoloff}, Dmitry},
        title = "{Galactic Magnetism: Recent Developments and Perspectives}",
      journal = {\araa},
         year = "1996",
        month = "Jan",
       volume = {34},
        pages = {155-206},
          doi = {10.1146/annurev.astro.34.1.155},
       adsurl = {https://ui.adsabs.harvard.edu/abs/1996ARA&A..34..155B}
}

@ARTICLE{2008Chyzy,
       author = {{Chy{\.z}y}, Krzysztof T. and {Buta}, Ronald J.},
        title = "{Discovery of a Strong Spiral Magnetic Field Crossing the Inner Pseudoring of NGC 4736}",
      journal = {\apjl},
         year = "2008",
        month = "Apr",
       volume = {677},
       number = {1},
        pages = {L17},
          doi = {10.1086/587958},
archivePrefix = {arXiv},
       eprint = {0803.0555},
 primaryClass = {astro-ph},
       adsurl = {https://ui.adsabs.harvard.edu/abs/2008ApJ...677L..17C}
}

@ARTICLE{1995_navarro_og,
       author = {{Navarro}, Julio F. and {Frenk}, Carlos S. and {White}, Simon D.~M.},
        title = "{Simulations of X-ray clusters}",
      journal = {\mnras},
         year = 1995,
        month = aug,
       volume = {275},
       number = {3},
        pages = {720-740},
          doi = {10.1093/mnras/275.3.720},
archivePrefix = {arXiv},
       eprint = {astro-ph/9408069},
 primaryClass = {astro-ph},
       adsurl = {https://ui.adsabs.harvard.edu/abs/1995MNRAS.275..720N}
}

@ARTICLE{1996_navarro_cored,
       author = {{Navarro}, Julio F. and {Frenk}, Carlos S. and {White}, Simon D.~M.},
        title = "{The Structure of Cold Dark Matter Halos}",
      journal = {\apj},
         year = 1996,
        month = may,
       volume = {462},
        pages = {563},
          doi = {10.1086/177173},
archivePrefix = {arXiv},
       eprint = {astro-ph/9508025},
 primaryClass = {astro-ph},
       adsurl = {https://ui.adsabs.harvard.edu/abs/1996ApJ...462..563N}
}

@BOOK{2019CFN,
       author = {{Cimatti}, Andrea and {Fraternali}, Filippo and {Nipoti}, Carlo},
        title = "{Introduction to galaxy formation and evolution: from primordial gas to present-day galaxies}",
    publisher = {Cambridge University Press},
         year = 2019,
       adsurl = {https://ui.adsabs.harvard.edu/abs/2019igfe.book.....C}
}

@article{crowther,
    author = {Crowther, Paul A. and Schnurr, Olivier and Hirschi, Raphael and Yusof, Norhasliza and Parker, Richard J. and Goodwin, Simon P. and Kassim, Hasan Abu},
    title = {The R136 star cluster hosts several stars whose individual masses greatly exceed the accepted 150 M⊙ stellar mass limit},
    journal = {\mnras},
    volume = {408},
    number = {2},
    pages = {731-751},
    year = {2010},
    month = {10},
    issn = {0035-8711},
    doi = {10.1111/j.1365-2966.2010.17167.x},
    url = {https://doi.org/10.1111/j.1365-2966.2010.17167.x},
    eprint = {https://academic.oup.com/mnras/article-pdf/408/2/731/18432602/mnras0408-0731.pdf},
}

@ARTICLE{yusof2010lifedeathmassivestars,
       author = {{Yusof}, Norhasliza and {Abu Kassim}, Hasan and {Hirschi}, Raphael and {Crowther}, Paul. A. and {Schnurr}, Olivier and {Parker}, Richard J. and {Goodwin}, Simon P.},
        title = "{Life and Death of Very Massive Stars}",
      journal = {arXiv e-prints},
         year = 2010,
        month = dec,
          eid = {arXiv:1012.3649},
        pages = {arXiv:1012.3649},
          doi = {10.48550/arXiv.1012.3649},
archivePrefix = {arXiv},
       eprint = {1012.3649},
 primaryClass = {astro-ph.SR},
       adsurl = {https://ui.adsabs.harvard.edu/abs/2010arXiv1012.3649Y}
}

@ARTICLE{2024Siljeg,
       author = {{{\v{S}}iljeg}, B. and {Adams}, E.~A.~K. and {Fraternali}, F. and {Hess}, K.~M. and {Oosterloo}, T.~A. and {Marasco}, A. and {Adebahr}, B. and {D{\'e}nes}, H. and {Garrido}, J. and {Lucero}, D.~M. and {Mancera Pi{\~n}a}, P.~E. and {Moss}, V.~A. and {Parra-Roy{\'o}n}, M. and {Ponomareva}, A.~A. and {S{\'a}nchez-Exp{\'o}sito}, S. and {van der Hulst}, J.~M.},
        title = "{Photometry and kinematics of dwarf galaxies from the Apertif H I survey}",
      journal = {\aap},
         year = 2024,
        month = dec,
       volume = {692},
          eid = {A217},
        pages = {A217},
          doi = {10.1051/0004-6361/202449923},
archivePrefix = {arXiv},
       eprint = {2409.18825},
 primaryClass = {astro-ph.GA},
       adsurl = {https://ui.adsabs.harvard.edu/abs/2024A&A...692A.217S}
}

@ARTICLE{2025ManceraPina,
       author = {{Mancera Pi{\~n}a}, Pavel E. and {Read}, Justin I. and {Kim}, Stacy and {Marasco}, Antonino and {Benavides}, Jos{\'e} A. and {Glowacki}, Marcin and {Pezzulli}, Gabriele and {Lagos}, Claudia del P.},
        title = "{The galaxy-halo connection of disc galaxies over six orders of magnitude in stellar mass}",
      journal = {\aap},
         year = 2025,
        month = jul,
       volume = {699},
          eid = {A311},
        pages = {A311},
          doi = {10.1051/0004-6361/202554381},
archivePrefix = {arXiv},
       eprint = {2505.22727},
 primaryClass = {astro-ph.GA},
       adsurl = {https://ui.adsabs.harvard.edu/abs/2025A&A...699A.311M}
}

@ARTICLE{2017Marasco,
   author = {{Marasco}, A. and {Fraternali}, F. and {van der Hulst}, J.~M. and 
	{Oosterloo}, T.},
    title = "{Distribution and kinematics of atomic and molecular gas inside the solar circle}",
  journal = {\aap},
archivePrefix = "arXiv",
   eprint = {1707.00743},
     year = 2017,
    month = nov,
   volume = 607,
      eid = {A106},
    pages = {A106},
      doi = {10.1051/0004-6361/201731054},
   adsurl = {http://adsabs.harvard.edu/abs/2017A%26A...607A.106M}
}

@ARTICLE{2022ManceraPina_b,
       author = {{Mancera Pi{\~n}a}, Pavel E. and {Fraternali}, Filippo and {Oosterloo}, Tom and {Adams}, Elizabeth A.~K. and {di Teodoro}, Enrico and {Bacchini}, Cecilia and {Iorio}, Giuliano},
        title = "{The impact of gas disc flaring on rotation curve decomposition and revisiting baryonic and dark matter relations for nearby galaxies}",
      journal = {\mnras},
         year = 2022,
        month = aug,
       volume = {514},
       number = {3},
        pages = {3329-3348},
          doi = {10.1093/mnras/stac1508},
archivePrefix = {arXiv},
       eprint = {2205.12977},
 primaryClass = {astro-ph.GA},
       adsurl = {https://ui.adsabs.harvard.edu/abs/2022MNRAS.514.3329M}
}

@article{Beattie_2026,
doi = {10.3847/2041-8213/ae6eed},
url = {https://doi.org/10.3847/2041-8213/ae6eed},
year = {2026},
month = {jun},
publisher = {The American Astronomical Society},
volume = {1004},
number = {1},
pages = {L9},
author = {Beattie, James R.},
title = {Supernovae Drive Large-scale, Incompressible Turbulence through Small-scale Instabilities},
journal = {\apjl}
}

@ARTICLE{Vishniac1983,
       author = {{Vishniac}, E.~T.},
        title = "{The dynamic and gravitational instabilities of spherical shocks}",
      journal = {\apj},
         year = 1983,
        month = nov,
       volume = {274},
        pages = {152-167},
          doi = {10.1086/161433},
       adsurl = {https://ui.adsabs.harvard.edu/abs/1983ApJ...274..152V}
}

@article{Facchini2026,
	author = {{Facchini}, P. and {Grebel}, E. K. and {Pasquali}, A. and {Sabbi}, E. and {Meena}, B. and {Bajaj}, V. and {Gallagher III}, J. S. and {Elmegreen}, B. G. and {Bianchi}, L. and {Adamo}, A. and {Calzetti}, D. and {Cignoni}, M. and {Crowther}, P. A. and {Eldridge}, J. J. and {Gennaro}, M. and {Klessen}, R. S. and {Smith}, L. J. and {Wofford}, A. and {Zeidler}, P.},
	title = {Isolated massive star candidates in NGC 4242 with the Galaxy UV Legacy Project},
	DOI= "10.1051/0004-6361/202555780",
	url= "https://doi.org/10.1051/0004-6361/202555780",
	journal = {\aap},
	year = 2026,
	volume = 705,
	pages = "A168",
}

@article{Sarbadhicary2026,
doi = {10.3847/1538-4357/ae3f33},
url = {https://doi.org/10.3847/1538-4357/ae3f33},
year = {2026},
month = {mar},
publisher = {The American Astronomical Society},
volume = {1000},
number = {1},
pages = {70},
author = {Sarbadhicary, Sumit K. and Wagner, Jordan and Koch, Eric W. and Mayker Chen, Ness and Leroy, Adam K. and Lahén, Natalia and Rosolowsky, Erik and Neugent, Kathryn F. and Kim, Chang-Goo and Chomiuk, Laura and Dalcanton, Julianne J. and Lopez, Laura A. and Pingel, Nickolas M. and Indebetouw, Remy and Williams, Thomas G. and Tarantino, Elizabeth and Donovan Meyer, Jennifer and Skillman, Evan D. and Smercina, Adam and Kepley, Amanda A. and Murphy, Eric J. and Strader, Jay and Wong, Tony and Stanimirović, Snežana and Villanueva, Vicente and Walter, Fabian and Ott, Juergen and Darling, Jeremy and Roman-Duval, Julia and Murray, Claire E.},
title = {Where Do Stars Explode in the ISM?—The Distribution of Dense Gas around Evolved Massive Stars in M33},
journal = {\apj}
}

@ARTICLE{FraternaliTomassetti2012,
       author = {{Fraternali}, F. and {Tomassetti}, M.},
        title = "{Estimating gas accretion in disc galaxies using the Kennicutt-Schmidt law}",
      journal = {\mnras},
         year = 2012,
        month = nov,
       volume = {426},
       number = {3},
        pages = {2166-2177},
          doi = {10.1111/j.1365-2966.2012.21650.x},
       eprint = {1207.0093},
       adsurl = {https://ui.adsabs.harvard.edu/abs/2012MNRAS.426.2166F},
}

@ARTICLE{KennicuttEvans2012,
       author = {{Kennicutt}, Robert C. and {Evans}, Neal J.},
        title = "{Star Formation in the Milky Way and Nearby Galaxies}",
      journal = {\araa},
         year = 2012,
        month = sep,
       volume = {50},
        pages = {531-608},
          doi = {10.1146/annurev-astro-081811-125610},
       adsurl = {https://ui.adsabs.harvard.edu/abs/2012ARA&A..50..531K},
}

@ARTICLE{Elmegreen_2024,
       author = {{Elmegreen}, Bruce G.},
        title = "{Feedback and Galaxy Dynamics: A Study of Turbulence and Star Formation in 34 Galaxies Using the PHANGS Survey}",
      journal = {\apj},
         year = 2024,
        month = may,
       volume = {966},
       number = {2},
          eid = {233},
        pages = {233},
       eprint = {2403.12927},
       adsurl = {https://ui.adsabs.harvard.edu/abs/2024ApJ...966..233E},
}
	
	\begin{appendix}
	\onecolumn
	\section{Galaxy sample properties}
	Table~\ref{tab:galaxy_data} lists the main properties of the dwarf galaxies in our sample.

	\begin{table*}
		\centering
		\caption{Physical properties of the galaxy sample.}
		\label{tab:galaxy_data}
			\begin{tabular}{lcccccc}
				\hline
				Galaxy & $D \pm \Delta D$ [Mpc] & $M_V$ [mag] & $M_{\mathrm{HI}}$ [$10^7\,M_\odot$] & $i \pm \Delta i$ [$^\circ$] & $\log \text{SFR} \pm \ \log \Delta \text{SFR}\ [M_\odot \text{yr}^{-1} ]$ & $M_\star$ [$10^7 M_\odot$]\\
				\hline
				CVnIdwA   & $3.6 \pm 0.108$ & $-12.4$ & 4.5 $\pm$ 0.1  & $49.2 \pm 10.9$ & $-2.47 \pm 0.07$ & 0.410\\
				DDO 101   & $6.4 \pm 5.350$ & $-13.4$ & 1.8 $\pm$ 0.2  & $52.4 \pm 1.7$ & $-2.46 \pm 0.06$ & 6.543\\
				DDO 126   & $4.9 \pm 0.324$ & $-14.9$ & 12.8 $\pm$ 0.1 & $62.2 \pm 2.9$ & $-1.81 \pm 0.04$ & 1.605\\
				DDO 133   & $5.1 \pm 1.117$ & $-15.7$ & 10.0 $\pm$ 0.2 & $38.9 \pm 3.7$ & $-1.90 \pm 0.03$ & 3.042\\
				DDO 154   & $3.7 \pm 0.514$ & $-14.2$ & 25.1 $\pm$ 0.1 & $67.9 \pm 1.1$ & $-1.89\pm 0.04$ & 0.835\\
				DDO 168   & $4.3 \pm 0.258$ & $-15.0$ & 25.1 $\pm$ 0.1 & $62.0 \pm 2.6$ & $-1.67\pm 0.04$ & 5.865\\
				DDO 210   & $0.9 \pm 0.386$ & $-10.9$ & 0.2 $\pm$ 0.1  & $63.2 \pm 3.2$ & $-3.85\pm 0.07$ & 0.068\\
				DDO 47    & $5.2 \pm 0.937$ & $-15.5$ & 35.3 $\pm$ 0.6 & $70.0 \pm 1.7$ & $-1.61\pm 0.03$ & 9.4\\
				DDO 50    & $3.4 \pm 0.205$ & $-16.6$ & 59.5 $\pm$ 1.6 & $37.4 \pm 4.6$ & $-0.94\pm 0.03$ & 10.725\\
				DDO 52    & $10.3 \pm 1.048$ & $-15.4$ & 22.3 $\pm$ 0.1 & $33.1 \pm 2.9$ & $-1.84\pm 0.05$ & 5.273\\
				DDO 53    & $3.6 \pm 0.193$ & $-13.8$ & 4.7 $\pm$ 0.3  & $55.1 \pm 2.0$ & $-2.12\pm 0.06$ & 0.970\\
				DDO 87    & $7.4 \pm 0.170$ & $-15.0$ & 20.7 $\pm$ 1.0 & $37.0 \pm 7.3$ & $-1.93\pm 0.04$ & 3.306\\
				NGC 2366  & $3.4 \pm 0.168$ & $-17.4$ & 64.0 $\pm$ 1.2 & $67.0 \pm 4.2$ & $-0.96\pm 0.05$ & 6.954\\
				WLM       & $1.0 \pm 0.064$ & $-14.4$ & 5.6 $\pm$ 0.1  & $67.6 \pm 2.3$ & $-2.14\pm 0.05$ & 1.629\\
				\hline
			\end{tabular}
		\tablefoot{Listed are the distance, absolute $V$-band magnitude, HI mass, inclination angle, logarithm of the total star formation rate, and stellar mass, with their respective uncertainties. The total star formation rates are calculated in this work.}
		\tablebib{\citet{zhang2012_stellarmass}, \citetalias{Bacchini_2020_volumetric}, \citet{Iorio_2017_data}.}
	\end{table*}

	\section{Radial profiles of energy components}
	In this appendix, we show the radial profiles of the energy components for the individual galaxies. Figures~\ref{fig:Energycomponents-CNM-single}, \ref{fig:Energycomponents-WNM-single}, and~\ref{fig:Energycomponents-2phase-single} show the characteristic SN efficiency for dwarfs in the CNM, WNM, and two-phase cases, respectively. Figures~\ref{fig:EnergyComponents-2phase-global} and~\ref{fig:GlobalEfficiency_2phase_all} show the "universal" SN efficiency for dwarfs only and for dwarfs and spirals together, respectively. Table~\ref{tab:SingleEffs} lists the characteristic SN efficiencies of the individual dwarf galaxies.

	\begin{table}
		\setlength\extrarowheight{2pt}
		\centering
		\caption{Characteristic SN efficiency to sustain the HI turbulence for each dwarf galaxy in our sample.}
		\label{tab:SingleEffs}
		\begin{tabular}{lccc}
			\toprule
			\toprule
			Galaxy & \multicolumn{3}{c}{SN efficiency} \\
			\cmidrule(lr){2-4}
			& $\log \eta_{\mathrm{CNM}}$ & $\log\eta_{\mathrm{WNM}}$ & $\log \eta_{\mathrm{2ph}}$ \\
			\midrule
			DDO 154   & $-1.31^{+0.13}_{-0.09}$ & $-2.05^{+0.29}_{-0.22}$ & $-1.64^{+0.14}_{-0.11}$ \\
			DDO 168  & $-1.06^{+0.11}_{-0.09}$ &$-1.54^{+0.20}_{-0.17}$ & $-1.15^{+0.14}_{-0.13}$ \\
			DDO 101  & $-1.91^{+0.21}_{-0.16}$ & $-4.19^{+2.86}_{-0.41}$ & $-2.32^{+0.66}_{-0.40}$ \\
			NGC 2366  & $-1.43^{+0.11}_{-0.09}$ & $-1.78^{+0.16}_{-0.12}$ & $-1.58^{+0.17}_{-0.13}$ \\
			DDO 126  & $-1.44^{+0.07}_{-0.06}$ & $-1.99^{+0.15}_{-0.13}$ & $-1.56^{+0.13}_{-0.13}$ \\
			CVnIdwA  & $-1.29^{+0.15}_{-0.11}$ & $-2.40^{+0.53}_{-0.35}$ & $-1.64^{+0.42}_{-0.26}$ \\
			DDO 47  & $-1.52^{+0.09}_{-0.08}$ & $-3.26^{+1.42}_{-0.41}$ & $-1.96^{+0.55}_{-0.36}$ \\
			DDO 53  & $-1.82^{+0.09}_{-0.08}$ &$-3.89^{+1.95}_{-0.41}$ & $-2.33^{+0.73}_{-0.40}$ \\
			WLM  & $-1.47^{+0.09}_{-0.13}$ & $-3.98^{+1.83}_{-0.40}$& $-3.82^{+2.13}_{-0.41}$ \\
			DDO 87 & $-1.93^{+0.15}_{-0.12}$ & $-4.36^{+2.79}_{-0.41}$ & $-2.33^{+0.56}_{-0.40}$\\
			DDO 50 & $-2.02^{+0.12}_{-0.10}$ & $-3.24^{+0.78}_{-0.38}$ & $-2.22^{+0.25}_{-0.26}$ \\
			DDO 52 & $-1.70^{+0.07}_{-0.06}$ & $-3.79^{+1.83}_{-0.41}$ & $-2.29^{+0.90}_{-0.41}$\\
			DDO 133 & $-1.52^{+0.07}_{-0.06}$ & $-3.28^{+1.42}_{-0.41}$ & $-1.83^{+0.37}_{-0.36}$\\
			DDO 210 & $-1.43^{+0.07}_{-0.06}$ & $-3.95^{+2.34}_{-0.41}$ & $-1.86^{+0.56}_{-0.39}$ \\
			\midrule
			Median       & $-1.50^{+0.68}_{-0.18}$ & $-3.17^{+8.24}_{-0.01}$ & $-1.91^{+1.14}_{-0.04}$ \\
			\bottomrule
			\bottomrule
		\end{tabular}
		\tablefoot{Since the WNM case reaches very low values, efficiencies are reported in log-scale. Each value corresponds to the posterior median, while the uncertainties are calculated as the 16th and 84th percentiles.}
	\end{table}

	\begin{figure*}[ht]
		\centering    \includegraphics[width=1\linewidth]{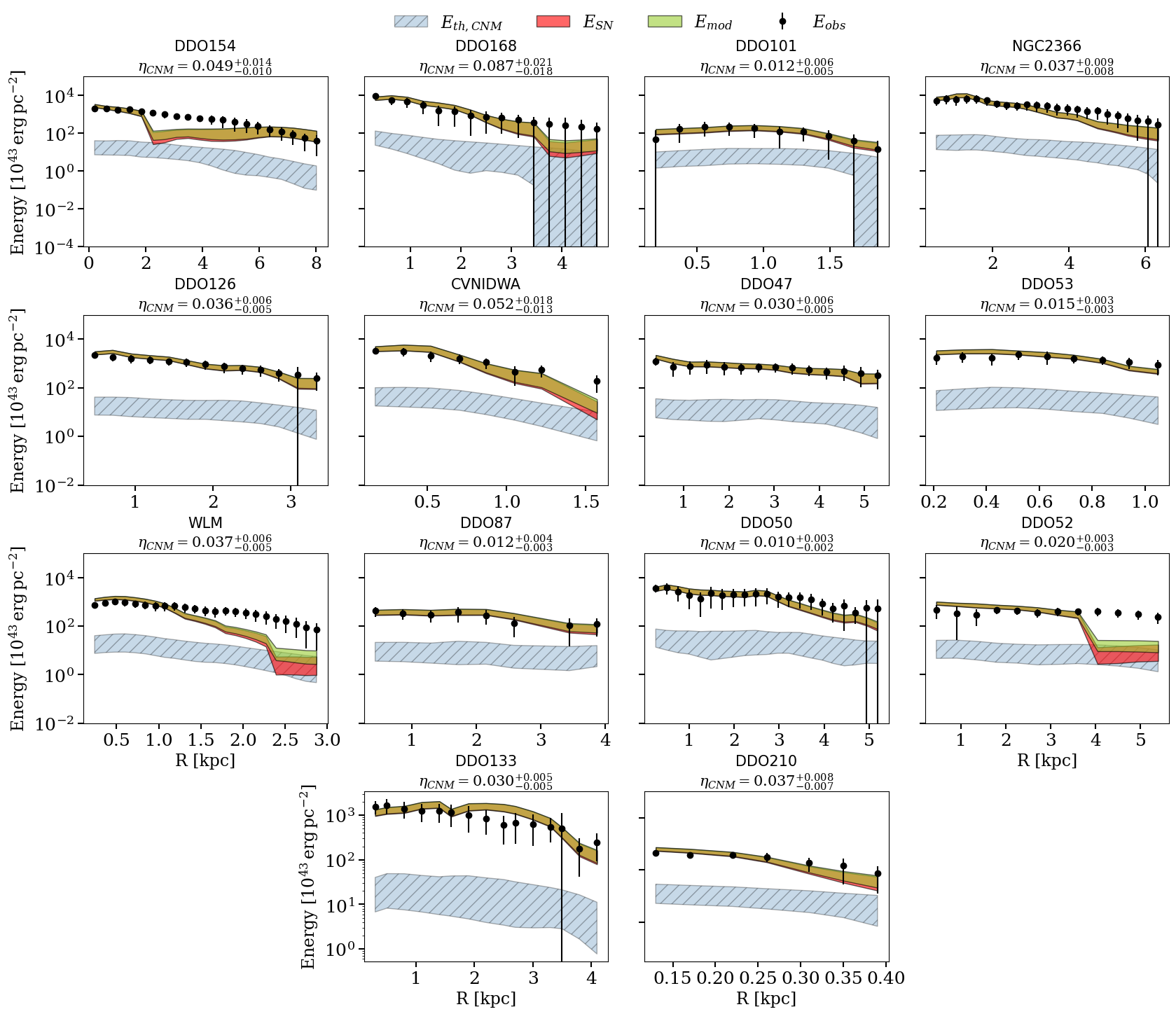}
		\caption{Radial profiles of the energies per unit area in the case assuming 100 \% CNM. The panels show, for each dwarf galaxy in our sample the observed energies (Eq. \ref{eq:obs_E}, black points),  the total model energy (Eq.~\ref{eq:E_mod}, green band), which includes both the turbulent energy from SNe (Eq.~\ref{eq:EturbSN}, red band) and the thermal energy (Eq.~\ref{eq:E_th}, hatched blue band). We note that some of the energy components may overlap in some cases.
		}
		\label{fig:Energycomponents-CNM-single}
	\end{figure*} 
	
	\begin{figure*}[ht]
		\centering
		\includegraphics[width=1\linewidth]{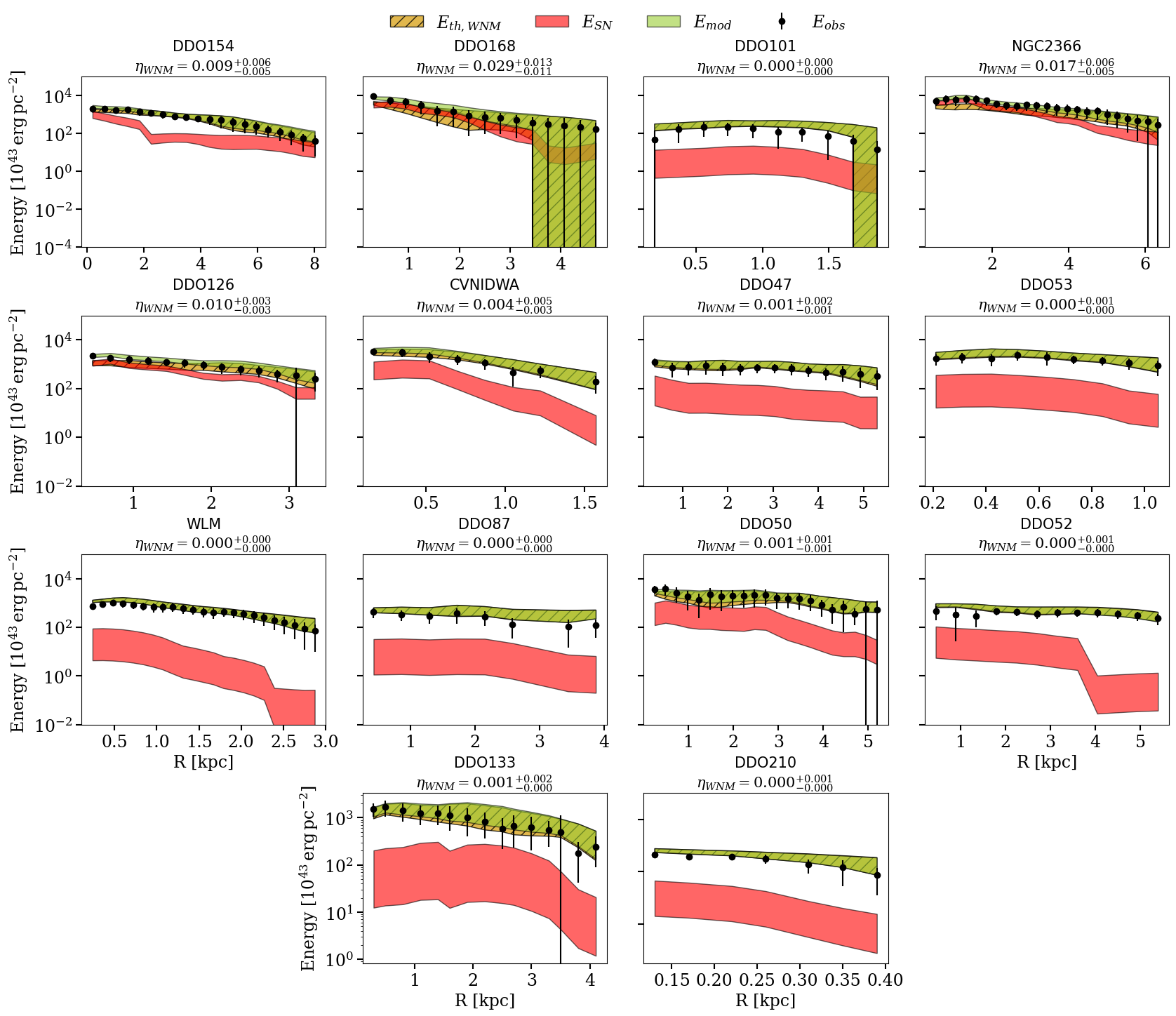}
		\caption{Same as Fig.~\ref{fig:Energycomponents-CNM-single}, but assuming 100 \% WNM. The thermal energy is now indicated by the hatched orange band.}
		\label{fig:Energycomponents-WNM-single}
	\end{figure*}
	
	\begin{figure*}[ht]
		\centering
		\includegraphics[width=1\linewidth]{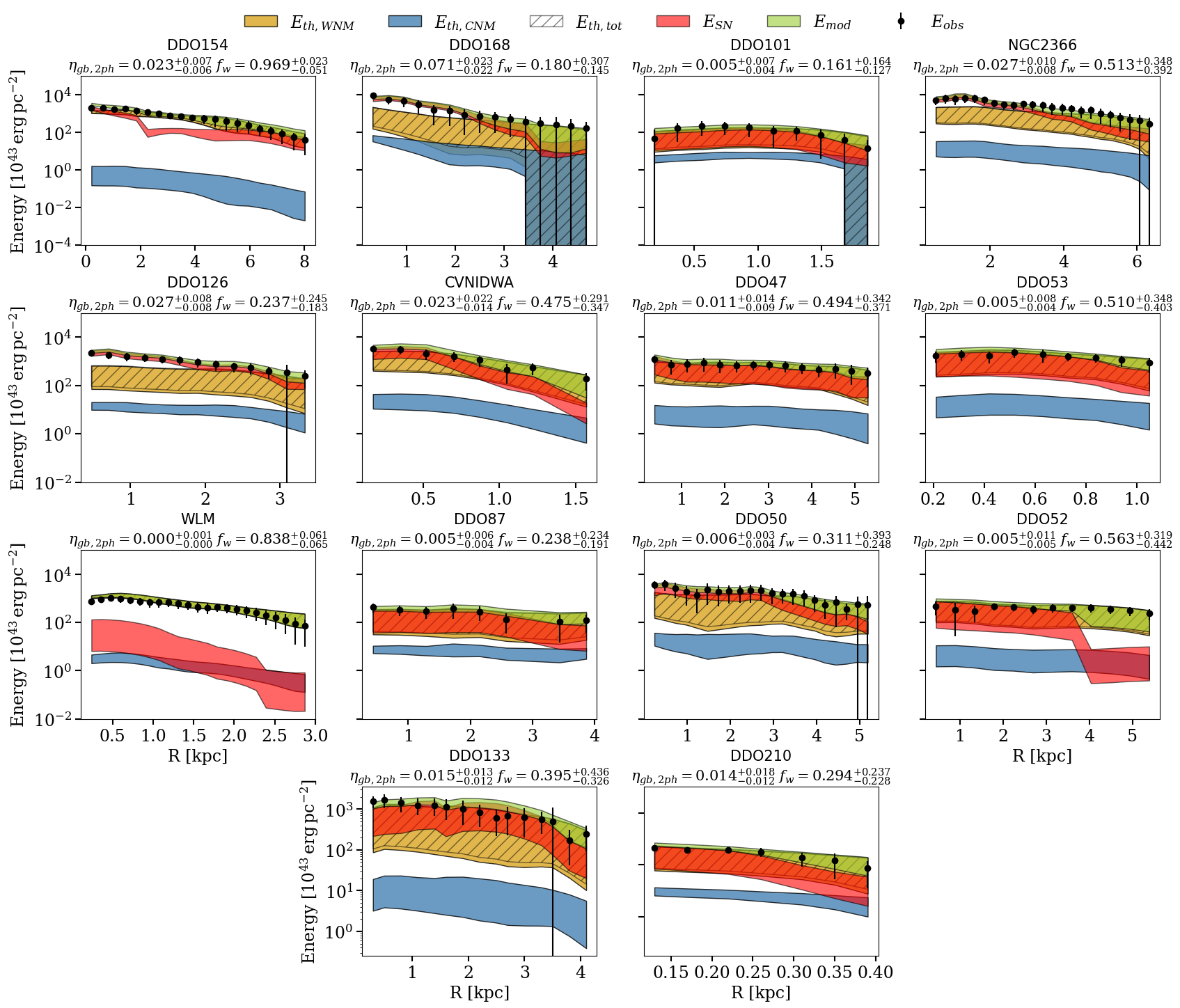}
		\caption{Same as in Fig.~\ref{fig:Energycomponents-CNM-single}, but for the two-phase gas case. 
			The blue, orange, and hatched bands show the thermal energies from CNM, WNM, and the whole neutral gas, respectively.}
		\label{fig:Energycomponents-2phase-single}
	\end{figure*}
	
	\begin{figure*}[ht]
		\centering
		\includegraphics[width=1\linewidth]{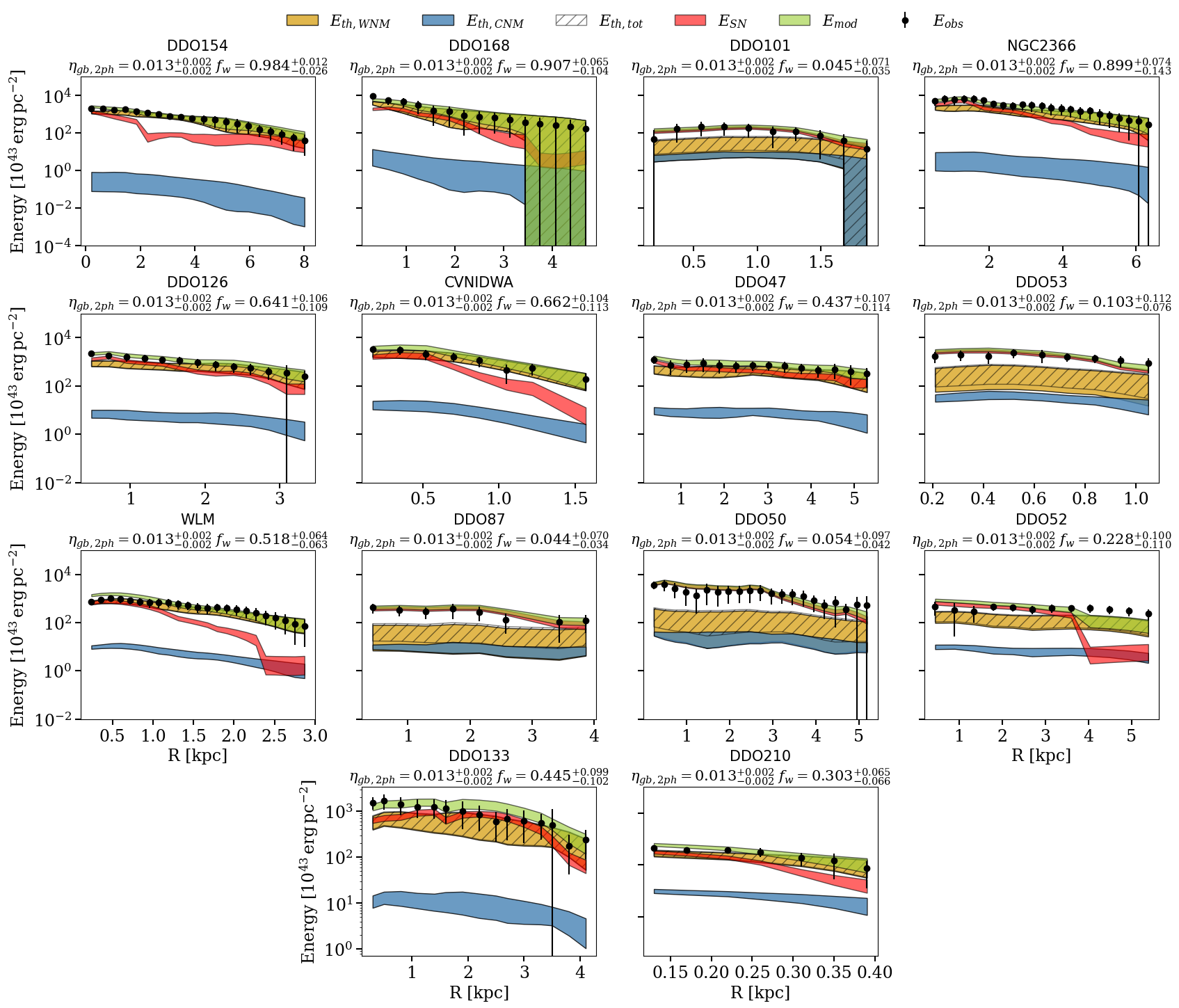}
		\caption{Same as Fig.~\ref{fig:Energycomponents-2phase-single} for the assumption of "universal" supernova efficiency ($\eta_\mathrm{gb,2ph}$) across all dwarf galaxies.}
		\label{fig:EnergyComponents-2phase-global}
	\end{figure*}
	
	\begin{figure*}[ht]
		\centering   \includegraphics[width=0.9\linewidth]{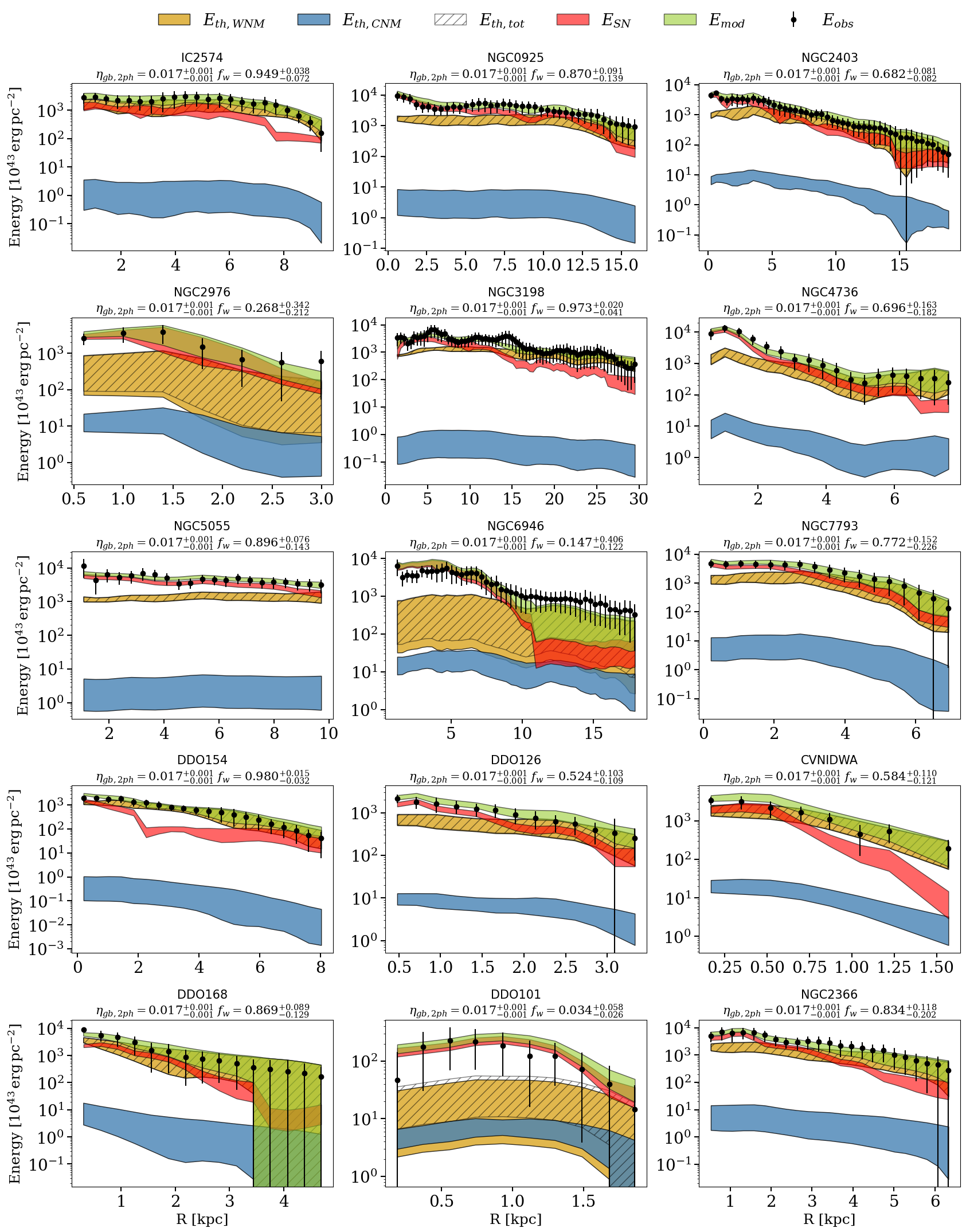}
	\caption{Same as Fig.~\ref{fig:Energycomponents-2phase-single} for the assumption of "universal" supernova efficiency ($\eta_\mathrm{gb,2ph}$) across all galaxies.}
	\label{fig:GlobalEfficiency_2phase_all}
\end{figure*}

\begin{figure*}[ht]\ContinuedFloat
	\hspace{0.225cm}\includegraphics[width=1\linewidth]{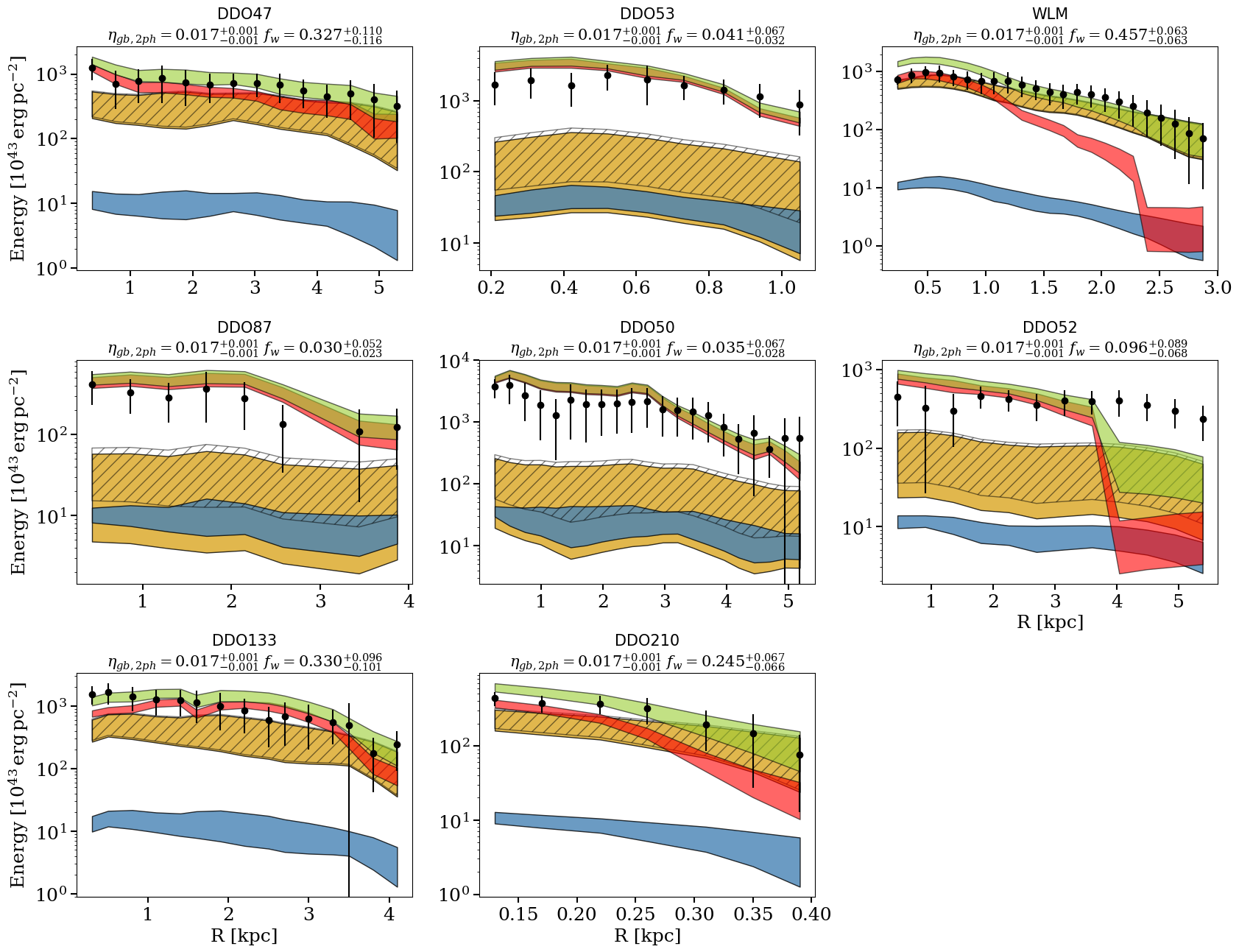}
	\caption{continued.}
\end{figure*}

\section{Secondary correlations}\label{app:radial_correlation}
Figures~\ref{fig:All_IN} and~\ref{fig:GlobalEfficiency_2phase} show strong correlations between the energy surface densities $E_\mathrm{SNe,max} \propto \Sigma_\mathrm{SFR} h / v_\mathrm{turb}$ (Eq.~\ref{eq:EturbSN}) and $E_\mathrm{turb} \propto \Sigma \sigma^2$ (Eq.~\ref{eq:obs_E}).
$\Sigma_\mathrm{SFR}$ and, to some degree, $\Sigma$ and $\sigma$ decline roughly exponentially with $R$, even though their scale lengths are not necessarily similar.
Therefore, the correlations in Figs.~\ref{fig:All_IN} and~\ref{fig:GlobalEfficiency_2phase} could be driven by the fact that the quantities on both axes have very similar radial gradients, rather than being a true physical connection via SN feedback. 

To assess this possibility, we remove from each galaxy the radial trend of $E_\mathrm{SNe,max}$ and $E_\mathrm{turb}$ and then verify whether the local deviations from that trend still correlate with each other. We do this in two ways, one parametric and the other non-parametric.
In the parametric approach, we fit to the profiles of each galaxy the function $E(R) = E_0 \, (1 + R/R_\mathrm{s})^{\alpha} \, \exp(-R/R_\mathrm{s})$ (Eq.~16 of \citealt{FraternaliTomassetti2012}), where $E_0$ is the central energy density, $\alpha$ is an exponent defining the tapering towards the centre, and $R_\mathrm{s}$ is a scale radius such that the peak of the distribution is located at $R = R_\mathrm{s} (\alpha - 1)$. 
This function is flexible enough to reproduce both the decline of the outer disc and the flattening seen in the inner regions of several dwarfs (Figs.~\ref{fig:Energycomponents-CNM-single}--\ref{fig:GlobalEfficiency_2phase_all}); the residual of each ring is the difference between the measured $\log E$ and the best-fit profile at that radius.
In the non-parametric approach, we divide the profiles of each galaxy into radial bins containing an equal number of points, with the number of bins set for each galaxy by the Freedman--Diaconis/Sturges rule; bins containing fewer than two points are discarded. 
We then subtract from each ring the median of $\log E$ of the bin it belongs to. This second method makes no assumption on the shape of the radial profile, so the residuals do not include the deviations of the true profile from an assumed functional form.
In both cases the residuals are the deviations, in dex, of each ring from the smooth radial trend of its own galaxy: they are dimensionless and carry no $10^{43}\,\mathrm{erg\,pc^{-2}}$ normalisation. 

Figure~\ref{fig:residuals} shows the residuals of $E_\mathrm{turb}$ versus those of $E_\mathrm{SNe,max}$ for the two prescriptions. 
The correlation persists in both cases: the Pearson correlation coefficient is $r = 0.46$ for the parametric method and $r = 0.65$ for the binned one and the probability $p$ that the correlation arises by chance is negligible. 
We have repeated the analysis for other prescriptions (e.g. a linear fit of $\log E$ against $\log (R/R_\mathrm{d})$, a second-order polynomial in $\log (R/R_\mathrm{d})$, and radial bins whose number is set by the square root of the number of rings). 
In all these tests, $r$ lies between $0.46$ and $0.80$, being largest for the most rigid prescriptions, which are also those that leave more of the radial trend in the residuals. 
Instead, the result is essentially insensitive to the minimum number of points required per bin. 
We conclude that a moderate-to-strong correlation between the residuals is present regardless of how the radial trend is removed, which indicates that the link between the SN energy and the turbulent energy is not simply a by-product of their common dependence on $R$.

\begin{figure*}[ht]
	\centering
	\includegraphics[width=1\linewidth]{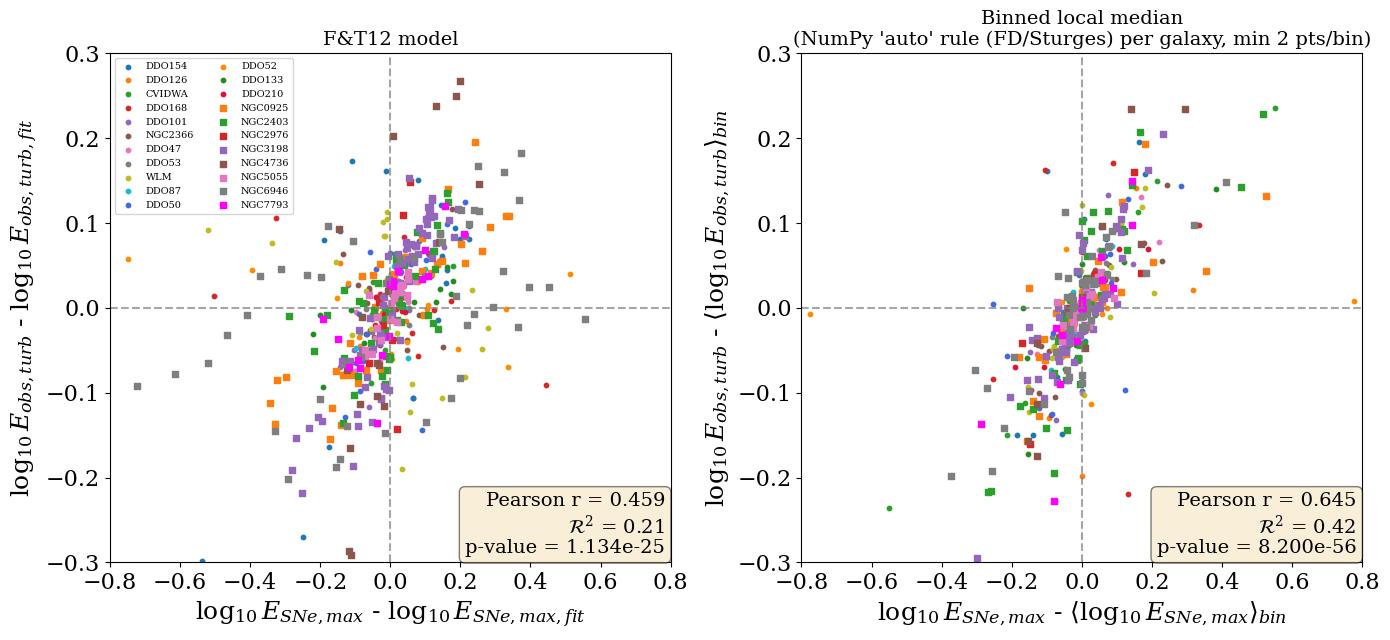}
	\caption{
			Residuals of the turbulent energy surface density versus residuals of the maximum SN energy surface density, for the two-phase case of Fig.~\ref{fig:GlobalEfficiency_2phase}. 
			The residuals are the deviations, in dex, of each ring from the radial trend of its own galaxy, and are therefore dimensionless. 
			Dashed lines mark null residuals.
			The Pearson correlation coefficient $r$, the coefficient of determination $\mathcal{R}^2$ and the $p$--value are reported in the bottom right corner of each panel. 
			Left panel: residuals from the best-fitting model based on Eq.~16 in \citet{FraternaliTomassetti2012}. 
			Right panel: residuals from the median of $\log E$ within radial bins containing an equal number of rings and at least two points. 
	}
	\label{fig:residuals}
\end{figure*}

Lastly, the correlations in Figs.~\ref{fig:All_IN} and~\ref{fig:GlobalEfficiency_2phase} should not be interpreted simply as a consequence of the correlation between the gas and SFR surface densities, that is the Kennicutt--Schmidt (KS) relation \citep{KennicuttEvans2012}. Our sample includes
dwarf irregulars and the outskirts of spirals, probing the low-density, HI-dominated regime where the classical KS relation breaks down (\citetalias{Bacchini_2020_volumetric}). If the correlation found here were solely inherited from the KS relation, one would expect a significantly larger scatter in Figs.~\ref{fig:All_IN} and~\ref{fig:GlobalEfficiency_2phase}; we find instead a remarkably tight relation. We also note that the quantities plotted there are not simply $\Sigma_\mathrm{SFR}$ and $\Sigma$: $E_\mathrm{turb}$ scales as $\Sigma \sigma^2$, while $E_\mathrm{SNe,max}$ carries the additional factor $h/v_\mathrm{turb}$ (the thickness crossing time), which is not constant across our sample (Fig.~\ref{fig:DissipationTime}) since it mostly depends on the galactic potential.

The correlations in Figs.~\ref{fig:All_IN} and~\ref{fig:GlobalEfficiency_2phase} have two readings: star formation follows the available gas (the usual reading of the KS relation), or turbulence follows star formation (the reading adopted here). Within the self-regulated cycle of Sect.~\ref{sec:self-regulating-cycle} these are not competing alternatives, as gravity sets the SFR from the gas volume density, SNe sustain the turbulence, and the turbulence regulates the disc thickness and hence the volume density itself. Consistently, the same data used here produce the volumetric star formation law (\citetalias{Bacchini_2019a}, \citetalias{2019b_bacchini}, \citetalias{Bacchini_2020_volumetric}). In any case, this study investigates the origin of the HI turbulence in the context of SN feedback, whatever mechanism sets the SFR. Furthermore, establishing whether SNe are the sole driver would require systems out of equilibrium \citep[see][Sect.~4.2]{Elmegreen_2024}, which is why we claim consistency rather than causality.

\end{appendix}

\end{document}